/% !TeX program = pdflatex

\documentclass[fleqn,usenatbib,useAMS]{mnras}

\usepackage{newtxtext,newtxmath}
\usepackage[T1]{fontenc}
\usepackage{ulem}
\usepackage[dvipsnames]{xcolor}
\defcitealias{Zhu_2023_paperI}{Paper I}
\usepackage{scalerel,tikz}
\usetikzlibrary{svg.path}
\definecolor{orcidlogocol}{HTML}{A6CE39}
\tikzset{orcidlogo/.pic={
		\fill[orcidlogocol] svg{M256,128c0,70.7-57.3,128-128,128C57.3,256,0,198.7,0,128C0,57.3,57.3,0,128,0C198.7,0,256,57.3,256,128z};
		\fill[white] svg{M86.3,186.2H70.9V79.1h15.4v48.4V186.2z}
		svg{M108.9,79.1h41.6c39.6,0,57,28.3,57,53.6c0,27.5-21.5,53.6-56.8,53.6h-41.8V79.1z M124.3,172.4h24.5c34.9,0,42.9-26.5,42.9-39.7c0-21.5-13.7-39.7-43.7-39.7h-23.7V172.4z}
		svg{M88.7,56.8c0,5.5-4.5,10.1-10.1,10.1c-5.6,0-10.1-4.6-10.1-10.1c0-5.6,4.5-10.1,10.1-10.1C84.2,46.7,88.7,51.3,88.7,56.8z};
}}
\newcommand\orcidicon[1]{\href{https://orcid.org/#1}{\mbox{\scalerel*{
				\begin{tikzpicture}[yscale=-1,transform shape]
					\pic{orcidlogo};
				\end{tikzpicture}
			}{|}}}}

\title[MaNGA DECaLS ggl]{MaNGA DynPop -- IV. Stacked total density profile of galaxy groups and clusters from combining dynamical models of integral-field stellar kinematics and galaxy-galaxy lensing}

\author[Chunxiang Wang et al.]{
	Chunxiang Wang$^{1,2,3}$\thanks{E-mail: \url{chunxiang_wang@sina.cn}},
	Ran Li$^{1,2,3}$\thanks{E-mail: \url{ranl@bao.ac.cn}},
	Kai Zhu\orcidicon{0000-0002-2583-2669}$^{1,2,3}$,
	Huanyuan Shan$^{4,10,11}$,
	Weiwei Xu\orcidicon{0000-0002-9587-6683}$^{1,2,3,5}$,\and 
	Michele Cappellari\orcidicon{0000-0002-1283-8420}$^{6}$,
	Liang Gao$^{1,2,3,7}$,
	Nan Li\orcidicon{0000-0001-6800-7389}$^{1,8}$,
	Shengdong Lu\orcidicon{0000-0002-6726-9499}$^{9,7}$,
	Shude Mao\orcidicon{0000-0001-8317-2788}$^{9,1}$,\and
	Ji Yao$^{4}$,
	Yushan Xie$^{3,4}$
	\\
	$^{1}$National Astronomical Observatories, Chinese Academy of Sciences, Beijing 100101, China\\
	$^{2}$Institute for Frontiers in Astronomy and Astrophysics, Beijing Normal University, Beijing 102206, China\\
	$^{3}$School of Astronomy and Space Science, University of Chinese Academy of Sciences, Beijing 100049, China\\
	$^{4}$Shanghai Astronomical Observatory (SHAO), Nandan Road 80, Shanghai 200030, China\\
	$^{5}$The Kavli Institute for Astronomy and Astrophysics, Peking University (KIAA-PKU), Beijing, China\\
	$^{6}$Sub-department of Astrophysics, Department of Physics, University of Oxford, Denys Wilkinson Building, Keble Road, Oxford, OX1 3RH, UK\\
	$^{7}$Institute for Computational Cosmology, Department of Physics, University of Durham, South Road, Durham, DH1 3LE, UK\\
	$^{8}$Key lab of Space Astronomy and Technology, National Astronomical Observatories, 20A Datun Road, Chaoyang District, Beijing 100012, China\\
	$^{9}$Department of Astronomy, Tsinghua University, Beijing 100084, China\\
	$^{10}$Key Laboratory of Radio Astronomy and Technology, Chinese Academy of Sciences, A20 Datun Road, Chaoyang District, Beijing, 100101, P. R. China\\
    $^{11}$University of Chinese Academy of Sciences, Beijing 100049, China\\}

\date{Submitted to MNRAS on 21 April 2023}

\pubyear{2023}

\begin{document}
	\label{firstpage}
	\pagerange{\pageref{firstpage}--\pageref{lastpage}}
	\maketitle
	
	\begin{abstract}
		We present the measurement of total and stellar/dark matter decomposed mass density profile around a sample of galaxy groups and clusters with dynamical masses derived from integral-field stellar kinematics from the MaNGA survey in \citetalias{Zhu_2023_paperI} and weak lensing derived from the DECaLS imaging survey. Combining the two data sets enables accurate measurement of the radial density distribution from several kpc to Mpc scales. Intriguingly, we find that the excess surface density derived from stellar kinematics in the inner region cannot be explained by simply adding an NFW dark matter halo extrapolated from lensing measurement at a larger scale to a stellar mass component derived from the NASA-Sloan Atlas (NSA) catalogue. We find that a good fit to both data sets requires a stellar mass normalization about 3 times higher than that derived from the NSA catalogue, which would require an unrealistically too-heavy initial mass function for stellar mass estimation. If we keep the stellar mass normalization to that of the NSA catalogue but allow a varying inner dark matter density profile, we obtain an asymptotic slope of $\gamma_{\rm gnfw}$= $1.82_{-0.25}^{+0.15}$ and $\gamma_{\rm gnfw}$=  $1.48_{-0.41}^{+0.20}$ for the group bin and the cluster bin respectively, significantly steeper than the NFW case. We also compare the total mass inner density slopes with those from TNG300 and find that the values from the simulation are lower than the observation by about $2\sigma$ level.   	
	\end{abstract}
	\begin{keywords}
		gravitational lensing:weak-galaxies:clusters:general-galaxies:statistics-dark matter
	\end{keywords}
	
	%%%%%%%%%%%%%%%%%%%%%%%%%%%%%%%%%%%%%%%%%%%%%%%%%%
	
	%%%%%%%%%%%%%%%%% BODY OF PAPER %%%%%%%%%%%%%%%%%%

	%%%%%%%%%%%%%%%%%%%%%%%%%%%%%%%%%%%%%%%%%%%%%%%%%%
	\section{Introduction}
	
	In the modern lambda cold dark matter ($\Lambda$CDM) cosmogony,  the structure forms hierarchically. Small dark matter haloes form early and then larger dark matter halos form through halo-halo merging and continuous dark matter accretion \citep{Frenk_and_White_2012}. N-body cosmological simulations show that in such an $\Lambda$CDM universe, the dark matter haloes obtain a self-similar universal structure of mass distribution \citep[e.g.][]{Navarro1996, Navarro_1997, Springel2008, Gao2011} that the spherically averaged dark matter mass profile follows $\rho(r) \propto r^{-1}$ at the inner part and $\rho(r) \propto r^{-3}$ at the outer part. In a real universe, galaxies form at the cente of dark matter haloes and co-evolve with dark matter haloes. The baryonic process during the galaxy formation and evolution can modify the mass distribution of the dark matter halo, especially at the inner part, through baryonic condensation and contraction \citep{Blumenthal1986, Gnedin2004, Gustafsson2006, Duffy2010, Schaller2015}, the expulsion of gas during the feedback process \citep[e.g.][]{NavarroEkeFrenk1996, Read2005, Pontzen2012}. The impact of these processes is complex and the net effect is not clear yet.  Accurate measurement of the total mass density profile, as well as the decomposed stellar and dark matter density profiles, can provide a unique tool to probe the galaxy formation process.

    The density profiles of real galaxies were measured mainly using dynamical modeling of the gas or stellar kinematics (see review by \citealt{Courteau2014}), or using gravitational lensing (see review by \citealt{Treu2010}).
 
    The presence of gas in a nearly circular motion in the equatorial plane of spiral galaxies made these the first targets for studies of rotation curves using either ionized gas at optical wavelengths \citep[e.g.][]{Rubin1980} or neutral \textsc{Hi} in the radio \citep[e.g.][]{Bosma1978}. These studies found that the rotation curves of spiral galaxies tend to be flat at large radius (beyond about 4 projected half-light radius $R_{\rm e}$) and provided one of the first convincing evidence for the presence of dark matter inside galaxies \citep{Faber1979}. The general flatness of rotation curves of spiral galaxies at large radius was later confirmed by numerous papers \citep[e.g.][]{Martinsson2013}.
 
    If one assumes the total density of galaxies to be well-approximated by power laws of the form $\rho_{\rm T}(r)\propto r^{-\gamma_{_{\rm T}}}$, there is a simple relation between the logarithmic slope $\gamma_{\rm vel}$ of the circular velocity and the one $\gamma_{_{\rm T}}$ of the density \citep[eq.~2.61]{Binney2008}		
    \begin{equation}
        \gamma_{_{\rm T}} = 2 - 2\gamma_{\rm vel}.
    \end{equation}
   This implies that the observed flat rotation curves ($\gamma_{\rm vel}\approx0$) of spiral galaxies suggest their total densities of spiral galaxies to be close to isothermal ($\gamma_{_{\rm T}}\approx2$) at large radius.

   Unlike spiral galaxies, early-type galaxies (ETGs) generally do not contain extended \textsc{Hi} discs. For ETGs the total density slopes were measured via strong gravitational lensing or stellar dynamics. The requirement for galaxies to act as a lens implies that samples of lenses galaxies tend to be ETGs with large masses and effective velocity dispersion $\sigma_{\rm e}$. For these lens ETGs the mean power-law slope out to a median radius of $R_{\rm e}/2$ was found to be $\langle\gamma_{\rm tot}\rangle=2.078$, or nearly isothermal, in the project Sloan Lens Advanced Camera for Surveys (SLACS) with a sample of 73 strong galaxy lenses \citep{Koopmans2009, Auger_2010b}. A similar value was reported in other strong lensing projects at higher redshift $z\sim 0.5$ \citep{Lirui2018}.

  Unlike strong gravitational lensing, stellar dynamical modeling can be applied to statistically-representative and larger samples of galaxies, without the need for the galaxies to be lenses. Moreover, one can sample different radius, not just close to the lens. However, stellar kinematics is challenging to measure out to large radius due to the low surface brightness. Various, early studies examined the density profiles of individual ETGs~\citep[e.g.][]{Weijmans_2008MNRAS.383.1343W, Weijmans_2009MNRAS.398..561W, Forestell_2010ApJ...716..370F, Morganti_2013MNRAS.431.3570M, Napolitano_2014MNRAS.439..659N}. \citet{Cappellari2015} conducted the first systematic investigation of the density profile of ETGs extending to large radius. They used the Jean Anisotropic Models (JAM, \citealt{Cappellari2008}) to analyze extended stellar kinematics for 14 massive ETGs from the SAGES Legacy Unifyinig Globulars and GalaxieS (SLUGGS) survey \citep{Brodie2014} out to a median radius of $4R_{\rm e}$ and found they are well described by power laws with the logarithmic slope of the total density profile slightly steeper than isothermal $\langle\gamma_{\rm tot}\rangle=2.19$ with a scatter of just $\sigma_\gamma=0.11$. This sample was later extended to 21 ETGs by \citet{Bellstedt2018} who found a very similar value $\langle\gamma_{\rm tot}\rangle=2.24$ with a scatter of $\sigma_\gamma=0.05$. This mean value and the small scatter were confirmed for a sample of 16 massive ETGs with \textsc{Hi} gas discs out to a median radius of $6R_{\rm e}$ by \citet{Serra2016}, who found a remarkably close mean slope $\langle\gamma_{\rm tot}\rangle=2.18$.  Subsequently, \citet{Poci2017} used JAM to model a much larger sample of 260 ETGs from the ATLAS$^{\rm 3D}$ survey \citep{Cappellari2011}, but only out to a median radius of $1R_{\rm e}$. They found that above a stellar dispersion $\lg(\sigma_{\rm e}/{\rm km\, s^{-1}})\approx2.1$, these galaxies have a mean total density slope of $\langle\gamma_{\rm tot}\rangle=2.193$, with an observed scatter of $\sigma_\gamma=0.17$, in excellent agreement with, the previous values, and consistently slightly larger than the lensing value. However, at lower $\sigma_{\rm e}$ a trend was discovered with the slope decreasing with $\gamma_{\rm tot}$. Moreover, the decrease of $\gamma_{\rm tot}$ was found to correlate better with $\sigma_{\rm e}$ than with stellar mass \citep[fig.~22c]{Cappellari2016}.

   The Sloan Digital Sky Survey (SDSS) Mapping Nearby Galaxies at Apache Point Observatory (MaNGA) survey \citep[SDSS-MaNGA,][]{Bundy2015} substantially expanded the size of galaxy samples with integral-field stellar kinematics and included both ETGs and spirals. Using these data \cite{liran_2019} derived the mass-weighted total density slope out to a median radius of $1.5R_{\rm e}$ with JAM for about 2K nearby galaxies, and again accurately reproduced the mean total inner density slope of $\langle\gamma_{\rm tot}\rangle=2.24$ for ETGs with $\sigma_{\rm e}\ga 100 \rm km s^{-1}$, below which the density slope decreases with $\sigma_{\rm e}$ as previously noted. However, their study was the first to model with a consistent method a large sample of both spiral galaxies and ETGs. They found a much larger variation and much clearer trend in the total slopes, with $\gamma_{\rm tot}$ decreasing, namely becoming more shallow, for spiral galaxies vs ETGs.

    We revisited the $\gamma_{\rm tot}$ trends in \citet{zhukai_2023_paper2}, using JAM modeling of a sample of about 6K galaxies with the best data quality, extracted from the final data release of the MaNGA survey. We again confirmed the nearly constant mean slope $\langle\gamma_{\rm tot}\rangle=2.20$ above $\lg(\sigma_{\rm e}/{\rm km\, s^{-1}})\approx2.2$ but also found a clear variation of $\gamma_{\rm tot}$ at fixed $\sigma_{\rm e}$, with the slopes decreasing for younger ages.
	
	Studies on galaxy groups and galaxy clusters\citep[e.g.][]{Newman_2015, Newman_2013} that combine stellar kinematics and weak gravitational lensing have shown that the total density slope may also be shallower than $2$, reaches $\sim 1.7$ for galaxy groups and $\sim 1.2$ for galaxy clusters.
	
	Unlike the total density distribution, the decomposed dark matter and baryonic density profiles usually cannot be measured reliably. The decomposition of the two density components is challenging when observational measurements are available only within about $1R_{\rm e}$, due to the model degeneracy between the total stellar mass and the dark matter inner density slope. It is important to have observational data from multiple scales,  which helps to obtain constraints on the mass of the dark matter halo, and shrinking the freedom of dark matter density profile \citep[e.g.][]{He2020, Newman_2015}. \citet{Sonnenfeld2012} derived dark matter density slope $\gamma=1.7 \pm 0.2$ for SDSSJ0946+1006, the `Jackpot' lens, by combining the stellar kinematics and the double Einstein ring data of the system, implying a scenario of dark matter contraction at the halo center. A value $\gamma=1.602\pm 0.079_{\rm syst}$, also consistent with halo contraction, was obtained for the Milky Way using APOGEE and Gaia data out to $5R_{\rm e}$ in what represents the most accurate determination for any galaxy, due to the availability of full six-dimensional stellar phase space information \citep{Nitschai2021}. On the other hand, \citet{Yang2020} derived decomposed mass models by combining stellar kinematics at inner one effective radius (1$R_{\rm e}$) and \textsc{Hi} kinematics within 5$R_{\rm e}$, and obtained a dark matter inner density slope $\gamma_{\rm dm}=0.6^{+0.3}_{-0.2}$ for NGC2974, a bright nearby ETG, which is much shallower than the standard Navarro-Frenk-White (NFW) value (1). \citet{Newman_2015} presented the dark matter inner density slope for a sample of galaxy groups and clusters, they found that the galaxy groups have a mean $\gamma_{\rm dm} \sim 1$, consistent with NFW, but the clusters have a mean $\gamma_{\rm dm}\sim 0.5$, significantly lower than the NFW prediction. Recently, \citet{Sartoris2020}, however, derive a $\gamma_{\rm dm}=0.99 \pm 0.04$ for Abell S1063. Overall, the measurement cases of decomposed density profiles are not enough and the results are controversial, thus independent observations are valuable to clarify the situation.
 
    In this work, we performed measurement of mass density profiles for a sample of galaxy groups and clusters by combining stellar kinematics from the integral field unit (IFU) data of the SDSS-MaNGA survey and the weak gravitational lensing measurement with Dark Energy Camera Legacy Survey (DECaLS; \citealt{Dey_2019}) shear catalogue. MaNGA data provides us the mass distribution information within the effective radius of the group/cluster central galaxy, and the stack galaxy-galaxy lensing measurement can constrain the mean density profile of the dark matter halo from $\sim 100$ kpc to a few Mpc. Combining the two data sets,  we derive the total mass density profile and decomposed stellar/dark matter profile for the selected galaxy groups and clusters.

	The structure of this paper is organized as follows. In Section ~\ref{sec:data}, we describe the data set. In Section~\ref{sec:methods}, we show the methods used in this work, including the dynamical model of the galaxies in Section~\ref{subsec:dynamicalmodel} and the gravitational lensing measurement in Section~\ref{subsec:galaxy-galaxy-lensing}. In Section~\ref{sec:result}, we present our results. Discussions and conclusions are shown in Section~\ref{sec:discussion_and_conclusion}. Throughout the paper, we adopt a flat $\Lambda$CDM cosmological model from the Planck 2015 results (\cite{Planck_Collaboration2016}, $\Omega_{\rm m}=0.3075$, $\rm H_{\rm 0}=67.74 \rm{km~s^{-1}~Mpc^{-1}}$).

	%%%%%%%%%%%%%%%%%%%%%%%%%%%%%%%%%%%%%%%%%%%%%%%%%%
	\section {Observational Data}
	\label{sec:data}
	
	In this project, we measure the lensing signal around galaxy groups and galaxy clusters in the overlapping region of the MaNGA and DECaLS survey, the latter of which provides the source catalogue of weak lensing measurements. We describe the data sets in this section. 
	
	\subsection{Lens galaxies} 
	
	We select our lens sample by matching the ETGs in MaNGA with the central galaxies of groups/clusters in the SDSS DR7 group catalogue~\citep[SDSSGC;][]{Yang2005,Yang_2007}.
	MaNGA~\citep{Bundy2015} is a multi-object IFU spectroscopy survey that makes use of the 2.5-meter Sloan Foundation Telescope and the Baryon Oscillation Spectroscopic Survey (BOSS) spectrograph \citep{Smee_2013}. The BOSS spectrograph provides continuous coverage between 3600 $\Angstrom$ and 10300 $\Angstrom$ with a spectral resolution R$\sim$2000. The full sample has been selected at low redshift ($0.01 < z < 0.15$) to follow a flat distribution of stellar mass across the range of  $\rm 10^9 \ M_{\odot} - 10^{12} \ M_{\odot}$. 
	
	We draw our lens sample from the final data release of MaNGA, which contains 10010 unique galaxies in the SDSS Data Release 17 (DR17, \citealt{Abdurrouf2022}). The MaNGA data cubes are obtained by the spectrophotometric calibration \citep{Yan2016} and the Data Reduction Pipeline (DRP; \citealt{Law2016}). For each data cube, the spaxels are Voronoi-binned \citep{Cappellari2003} to reach a target S/N $\sim$10. The MaNGA Data Analysis Pipeline (DAP; \citealp{Westfall2019}), which uses \textsc{ppxf} \citep{Cappellari2017} and a subset of MILES library \citep{Sanchez-Blazquez2006}, MILES-HC, extracts the stellar kinematics for each binned spectrum. The stellar kinematics from DAP products publicly available since SDSS Data Release 17\footnote{\url{https://www.sdss.org/dr17/manga/manga-data}} provide the spatial distribution of the projected stellar velocity and stellar velocity dispersion for each galaxy. \citet[hereafter Paper I]{Zhu_2023_paperI} used the MaNGA stellar kinematics to construct accurate dynamical models of these galaxies and we use the result from the models in this paper.
	
	The following criteria are used in the lens selection process,
   \begin{itemize}
      \item{$\Theta_{\rm M,S}<1.0$ arcsec.}
		\item{$\Delta z<0.01$.}
		\item{${\rm Qual} \geq 0$.}
   \end{itemize}

	Here, the $\Theta_{\rm M, S}$ is the angular separation between the MaNGA galaxy and the central galaxies in SDSSGC, where the central galaxies are defined as the galaxy with the largest stellar mass, and $\Delta z$ is the difference between the corresponding redshifts from the two catalogues. \citetalias{Zhu_2023_paperI} assigns a quality grade ($\rm Qual = -1, 0, 1, 2, 3$) to each MaNGA galaxy. The higher the grade is, the better the model fitting quality is. We discard those galaxies with $\rm Qual=-1$ whose kinematic/dynamic properties can not be trusted. For galaxies with ${\rm Qual} \geq 0$, we further require the ${\rm lg}\left[ M_{\rm T}(<R_{\rm e})_{\rm gnfw,sph} \right]$ and ${\rm lg}\left [ M_{\rm T}(<R_{\rm e})_{\rm gnfw,cyl} \right ]$ to be consistent within 1$\sigma$ ($\frac{0.13}{\sqrt 2 }$ dex, derived from \textsc{lts\_linefit} \footnote{Version 5.0.19, from \url{https://pypi.org/project/ltsfit/}} software by fitting galaxies with Qual=0 without clipping), where the $M_{\rm T}(<R_{\rm e})_{\rm gnfw,sph}$ and $M_{\rm T}(<R_{\rm e})_{\rm gnfw,cyl}$ are the total mass within a sphere of radius $R_{\rm e}$ measured by dynamical models assuming either a spherically-aligned "$\rm JAM_{sph}$"+ "gNFW" model or cylindrically-aligned velocity ellipsoid "$\rm JAM_{cyl}$"+ "gNFW" model, respectively.  We refer readers to \citetalias{Zhu_2023_paperI} for model details.
	
	In this work, we only use groups/clusters whose central galaxy is ETG defined in \citet{Dominguez-Sanchez2022}. We removed the late-type galaxies to suppress the mis-center effect~\citep{Gao2006MNRAS.373...65G,Leauthaud2010ApJ...709...97L,Shan2017,Wangcx_2018}. We further require the lens sample to overlap with the survey footprint of DECaLS, and we remove lenses with redshift $z<0.02$ to ensure an efficient measurement of galaxy-galaxy lensing.
	We divide these central galaxies into 3 bins according to their assigned halo mass $M_{\rm 200m}$ in the SDSSGC, where $M_{\rm 200m}$ is defined as the total mass enclosed in $R_{\rm 200m}$ within which the mean density is 200 times of the mean matter density of the universe at the redshift of the halo. The mass range  and the number of lenses in different bins are listed in Table \ref{tab:fitting_result}. 
	%   table1
\begin{table*}
\begin{center}
\caption{Posterior constraints of free parameters of different models. The first three columns, show the halo mass range, the number of lenses, and the data set and model. The following four columns show the fitting parameters, halo mass, concentration, stellar mass normalization,  dark matter fraction within the mean value of $R_{\rm e}$, and reduced $\chi^2/\nu$. The best-fit value represents the peak of the one-dimensional marginalized posterior distribution of one parameter. We determine a horizontal line that intersects the marginalized distribution of the parameter at two points, such that the probability between these two intersections sums up to 68\%. These two intersections correspond to the upper and lower limits of the $1\sigma$ interval.}
\begin{tabular}{llllllll}
 \hline  \hline
 Halo mass $M_{\rm 200m}$ range
 & Nlens
 & Signal and model 
 &${\rm lg(}M_{\rm 200}{\rm [M_{\odot}])}$
 & ${C_{\rm 200}}$
 & $\alpha_{\rm nsa}$
 & $f_{\rm dm}(<\left<R_{\rm e}\right>)$
 & $\chi^{2}/\nu$\\
 \hline
 $M_{\rm 200m}<10^{13}{\rm M_{\odot}}$
& 512
& $---$
& $---$
& $---$
& $---$
& $---$
& $---$\\
$10^{13}{\rm M_{\odot}} <M_{\rm 200m}<10^{14}{\rm M_{\odot}}$
& 422
& ggl only
& $13.19_{-0.16}^{+0.15}$
& $2.73_{-1.7}^{+3.8}$
& $---$
& $0.28_{-0.13}^{+0.2}$
& 0.33\\ 
$10^{13}{\rm M_{\odot}} <M_{\rm 200m}<10^{14}{\rm M_{\odot}}$
& 422
& ggl+dyn
& $13.07_{-0.14}^{+0.13}$
& $13.26_{-1.52}^{+3.02}$
& $---$
& $0.66_{-0.03}^{+0.02}$
& 0.88\\ 
$10^{13}{\rm M_{\odot}} <M_{\rm 200m}<10^{14}{\rm M_{\odot}}$
& 422
& ggl+dyn+free ml
& $13.18_{-0.17}^{+0.14}$
& $2.32_{-1.52}^{+3.78}$
& $3.34_{-0.73}^{+0.52}$
& $0.06_{-0.04}^{+0.14}$
& 0.36\\ 
$M_{\rm 200m}>10^{14}{\rm M_{\odot}}$
& 97
& ggl only
& $13.89_{-0.12}^{+0.1}$
& $4.54_{-1.66}^{+2.39}$
& $---$
& $0.6_{-0.11}^{+0.1}$
& 1.39\\ 
$M_{\rm 200m}>10^{14}{\rm M_{\odot}}$
& 97
& ggl+dyn
& $13.87_{-0.12}^{+0.09}$
& $7.01_{-1.49}^{+1.9}$
& $---$
& $0.68_{-0.05}^{+0.05}$
& 1.44\\ 
$M_{\rm 200m}>10^{14}{\rm M_{\odot}}$
& 97
& ggl+dyn+free ml
& $13.88_{-0.11}^{+0.1}$
& $4.53_{-1.71}^{+2.05}$
& $2.87_{-1.38}^{+1.12}$
& $0.23_{-0.11}^{+0.24}$
& 1.41\\

 \hline
 \hline
\end{tabular}
\label{tab:fitting_result}
\end{center}
\end{table*}
	
\begin{figure*}
		\centering
		\includegraphics[width=1.6\columnwidth]{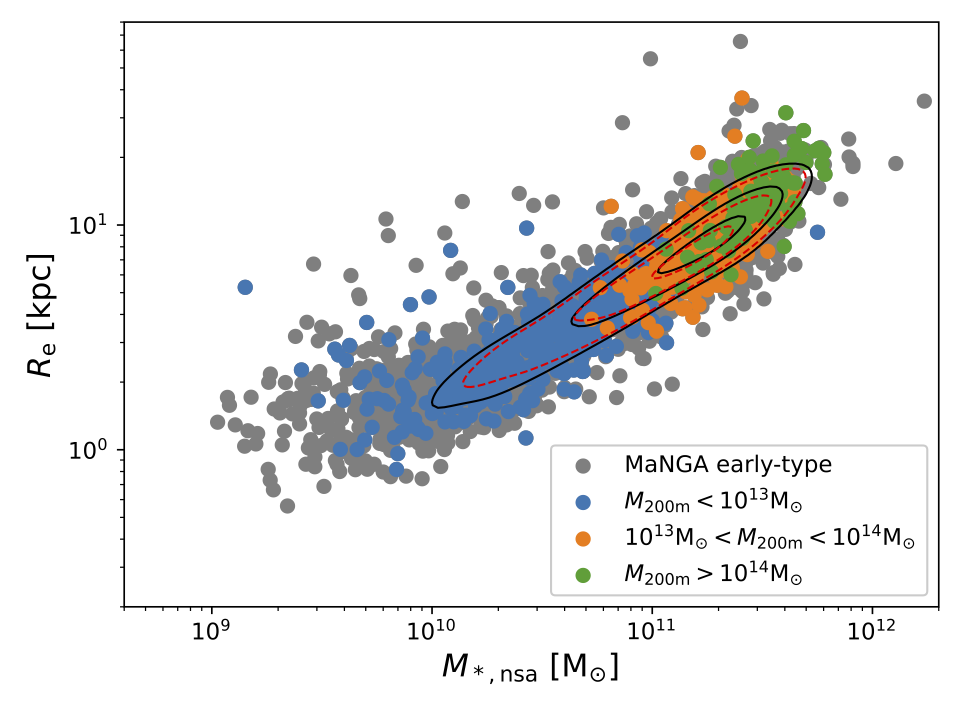}
		\caption{The relation between the NSA stellar mass and the effective radius $R_{\rm e}$ of the selected lens galaxies in different group masses bins. Colored dots show the sample with different masses. The full sample of MaNGA ETGs are plotted using  grey dots. Contours show the full sample of MaNGA ETGs (solid lines) and the combined there sub-samples used in this study (dashed lines) at 30\%, 60\%, and 90\% probability levels.}
		\label{fig:stellarMASS_vs_Re}
	\end{figure*}

	In Fig.~\ref{fig:stellarMASS_vs_Re}, we show the stellar mass and effective radius of our lens galaxies and the full sample of MaNGA ETGs, where the values are derived from the NASA-Sloan Atlas (NSA) catalogue\footnote{\url{http://nsatlas.org/}} \citep{Blanton2007, Blanton2011}. The $R_{\rm e}$ is the circularized effective radius which is calculated from the multi-Gaussian expansion \citep[MGE;][]{Emsellem1994, Cappellari2002} formalism of the galaxy $r$-band luminosity distribution. $R_{\rm e}$ is finally scaled by a factor of 1.35 to match the values determined from 2MASS \citep{Skrutskie2006} plus RC3 \citep{deVaucouleurs1991}. In Fig.~\ref{fig:stellarMASS_vs_Re}, the three sub-samples with increasing halo mass are represented by blue, orange, and green dots. The full sample of MaNGA ETGs is depicted by gray dots. Contours show the full sample of MaNGA ETGs (solid lines) and the combined there sub-samples used in this study (dashed lines) at 30\%, 60\%, and 90\% probability levels. One can see that there is only a slight difference between the contours of these two samples, indicating that the selected sources represent well the population of ETGs in the MaNGA full dataset.
	
\begin{figure}
		\centering
		\includegraphics[width=\columnwidth]{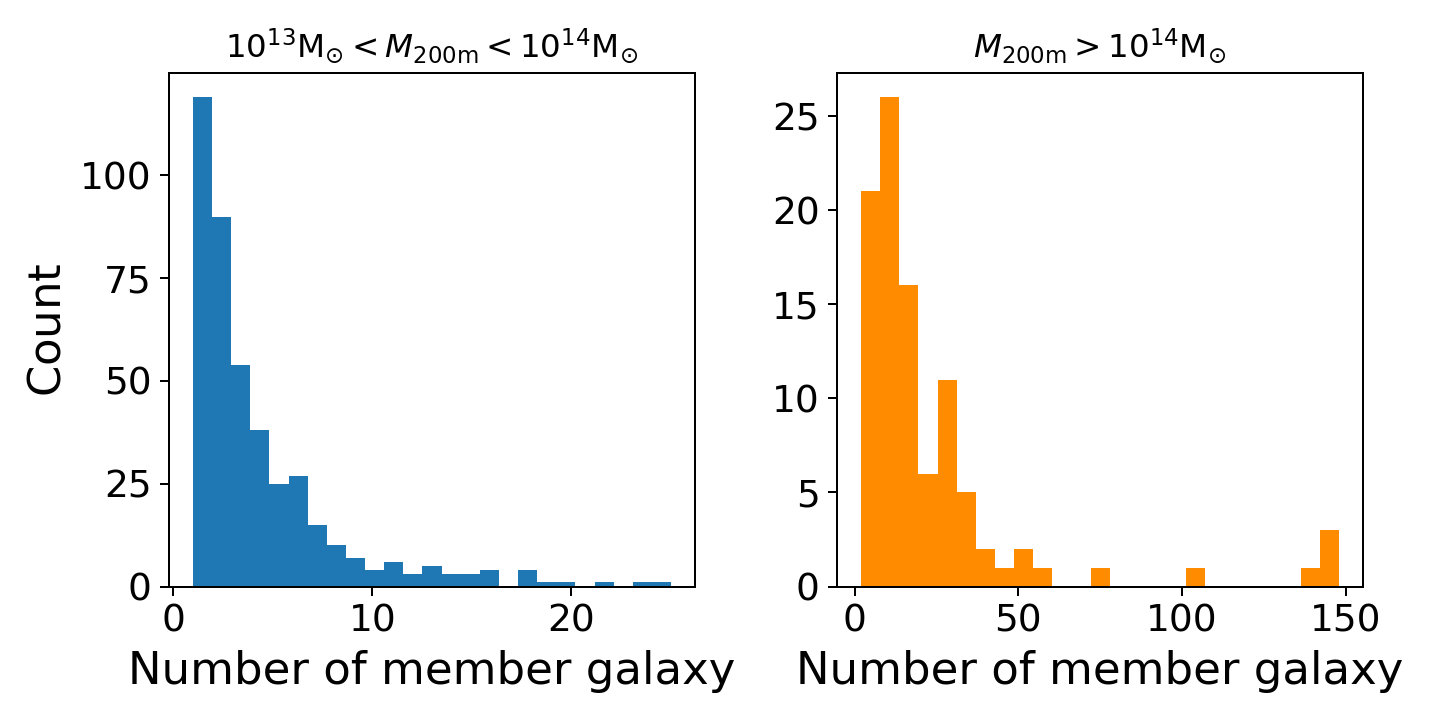}
		\caption{Distribution of the number of member galaxies in SDSSGC. We show that of the group bin and cluster bin in the left and right panels, respectively.}
		\label{fig:distribution_of_member_galaxy}
	\end{figure}

	In this work, we focus on the median and the high mass bins, which we will refer to as the group bin and the cluster bin respectively. The observational lensing error of the lowest mass bin is too large to derive meaningful results, as shown in the Appendix. Distributions of the number of member galaxies from SDSSGC in the group bin and the cluster bin are shown in Fig.~\ref{fig:distribution_of_member_galaxy}.


	\subsection{Source galaxies}
	\label{ssec:DECaLS_survey}
	In this work, we use the Dark Energy Camera Legacy Survey\footnote{\url{https://www.legacysurvey.org/}} \citep[DECaLS, see][]{Dey_2019} shear catalogue which has been utilized in multiple scientific studies~\citep[e.g.][]{Phriksee2020MNRAS.491.1643P, Xu_2021}. The source galaxies are taken from data release 8 (DR8) of DECaLS and the sky coverage of DECaLS DR8 is $\sim$9500 $\rm{deg^{2}}$ in $grz$ bands.
	
	The DECaLS DR8 data is processed by \textsc{tractor}~\citep{Meisner2017,Lang2016AJ....151...36L}. The morphologies of sources are divided into five types, including point sources (PSF), simple galaxies (SIMP, an exponential profile with affixed $0\farcs45$ effective radius and round profile), DeVaucouleurs (DEV, elliptical galaxies), Exponential (EXP, spiral galaxies), and Composite model (COMP, deVaucouleurs + exponential profile with the same source center)\footnote{\url{https://www.legacysurvey.org/dr8/description/}}. Sky-subtracted images are stacked in five different ways: one stack per band, one flat Spectral Energy Distribution (SED) stack of the $g$, $r$, $z$ bands,  and one red SED stack of all bands ($g-r=1$~mag and $r-z=1$~mag). Sources above the 6$\sigma$ detection limit in any stack are kept as candidates. Galaxy ellipticities (e1,e2) are estimated by a joint fitting image of $g$, $r$, and $z$ bands for  SIMP, DEV, EXP, and COMP galaxies. The multiplicative bias ($m$) and additive biases \citep[e.g.][]{Heymans2012MNRAS.427..146H, Miller2013} are modeled by calibrating with the image simulation \citep{Phriksee2020} and cross-matching  with external shear measurements 
	\citep{Phriksee2020,Yao2020,Zu2021}, including the Canada-France-Hawaii Telescope (CFHT) Stripe 82 \citep{Moraes2014}, Dark Energy Survey \citep{DES2016}, and Kilo-Degree Survey \citep{Hildebrandt2017} objects.

	The photo-$z$ of each source galaxy in DECaLS DR8 shear catalogue is taken from  \citet{Zou2019}, where the redshift of a target galaxy is derived with its k-nearest-neighbor in the SED space whose spectroscopic redshift is known. The photo-z is derived using 5 photometric bands: three optical bands, $g$, $r$, and $z$ from DECaLS DR8, and two infrared bands, W1, W2, from Wide-Field Infrared Survey Explorer. By comparing with a spectroscopic sample of 2.2 million galaxies, \citet{Zou2019} shows that the final photo-z catalogue has a redshift bias of $\Delta \overline{z}_{\rm norm}=2.4\times10^{-4}$, the accuracy of $\sigma_{\Delta z_{\rm norm}}=0.017$, and outlier rate of about 5.1\%.

	%%%%%%%%%%%%%%%%%%%%%%%%%%%%%%%%%%%%%%%%%%%%%%%%%%
	\section{Methods}
	\label{sec:methods}
	\subsection{Dynamical Model}
	\label{subsec:dynamicalmodel}
	
	In this paper, we make use of dynamical models lens galaxies derived in \citetalias{Zhu_2023_paperI}, where the JAM~\citep{Cappellari2008, Cappellari2020} is performed for the final SDSS-MaNGA data release. The JAM method has been applied to the mock stellar kinematic data from cosmological simulations and is demonstrated to be robust in recovering the total mass profile \citep{Lablanche2012, Lihongyu2016}. Using the JAM method, \citetalias{Zhu_2023_paperI} predicts second velocity moments maps with an assumed parametric mass distribution model and fits them to the observed one extracted from MaNGA.
	
	To investigate the systematics introduced by different theoretical assumptions and different parametric forms, \citetalias{Zhu_2023_paperI} uses eight mass models to fit the observed stellar kinematics. They adopt two extreme assumptions on the orientation of velocity ellipsoid, i.e. $\rm JAM_{cyl}$ (cylindrically-aligned) and $\rm JAM_{sph}$ (spherically-aligned), and for each orientation case, they adopt four parametric mass distributions, including one mass-follow-light mass model and three stellar-dark-matter two-component models.
	
	In this work, we do not fit stellar kinematics and weak lensing signal simultaneously, mainly because 
	the dynamical systems cannot be stacked linearly as the galaxy-galaxy lensing signal. Instead, we use the average value of $M_{\rm dyn}(<R_0)$, the JAM derived 3D total mass enclosed in a radius of $R_0$, for different mass ranges, as the observational constraints. We note that an alternative approach would consist of constraining the outer density to follow the average values for the halo mass when fitting the dynamical models of individual galaxies with JAM.
	
	The primary sample in SDSS-MaNGA has IFU data within $1.5R_e$, within which the dynamical total mass estimation are most reliable, thus we set $R_0$ to be the $1.5\left<R_e\right>$ of the central galaxies. $R_{\rm e}$ is the effective radius. For the group and the cluster bins, $R_0$ are $13.57$ and $20.10$ kpc respectively.

	For each halo mass bin, we calculate the $\left< M_{\rm dyn}(<R_0)\right>$ using the  "$\rm JAM_{cyl}$ "+ "gNFW" model of \citetalias{Zhu_2023_paperI} which is the most flexible model in \citetalias{Zhu_2023_paperI}. We estimate $\sigma_{M,\rm dyn}$, the uncertainties of $\left <M_{\rm dyn}(<R_0)\right>$ as follows. First, we use the bootstrap method to estimate the statistical uncertainties $\sigma_{\rm stat}$ using the total mass estimated by "$\rm JAM_{cyl}$ "+ "gNFW" model. Then, we estimate the systematic error $\sigma_{\rm sys}$ of $\left<M_{\rm dyn}\left(<R_0\right)\right>$ induced by imperfect model assumptions as $\sigma_{\rm sys}=\sqrt{\sum{\frac{\sigma_i^2}{N}}}$, where $\sigma_i$ is standard deviation of $M_{\rm dyn}(<R_0)$ among 8 different mass models of \citetalias{Zhu_2023_paperI} for the $i$th galaxy.  Finally, we have  $\sigma_{M,\rm dyn}^2=\sigma_{\rm stat}^2+\sigma_{\rm sys}^2$. 
	
	\subsection{Galaxy-Galaxy Lensing}
	\label{subsec:galaxy-galaxy-lensing}
    We measure the stacked galaxy-galaxy lensing signal in 10 logarithmic radial bins from 0.1 to 10 Mpc, where the excess surface density, $\Delta \Sigma (R)$ is derived as
	\begin{equation}
		\Delta \Sigma (R) =\overline{\Sigma}(<R)-\overline{\Sigma}(R)=\frac{\sum_{\rm ls}\omega_{\rm ls}\gamma_{\rm t}^{\rm ls}\Sigma_{\rm crit}}{\sum_{\rm ls}\omega_{\rm ls}} \,,
	\end{equation}
	where \noindent $\overline{\Sigma}(<R)$ is the mean density within the radius $R$, the $\overline{\Sigma}(R)$ is the azimuthally averaged surface density at radius $R$~\citep[e.g.][]{Miralda-Escude1991ApJ...370....1M,Wilson2001ApJ...555..572W,Leauthaud2010ApJ...709...97L}, $\gamma_{\rm t}^{\rm ls}$ is the tangential shear, and
	\begin{equation}
		\omega_{\rm ls}=\omega_{\rm n}\Sigma_{\rm crit}^{-2}\,,
	\end{equation}
	where the critical surface density  $\Sigma_{\rm crit}$ can be written as
	\begin{equation}
		\Sigma_{\rm crit}=\frac{c^2}{4 \rm \pi \rm G}\frac{D_{\rm s}}{D_{\rm l}D_{\rm ls}} \,,
	\end{equation}
	where $D_{\rm s}$, $D_{\rm l}$, and $D_{\rm ls}$ are respectively the angular diameter distance between the observer and the source, the observer and the lens, and the source and lens, and $c$ is the constant of light speed in the vacuum. The weight factor $\omega_{\rm n}$ is introduced to account for intrinsic scatter in ellipticity and shape measurement error of each source galaxy \citep{Miller2007, Miller2013}, defined as $\omega_{\rm n}=1/(\sigma^2_{\epsilon}+\sigma^2_{\rm e})$, where  $\sigma_{\epsilon}=0.27$ is the intrinsic ellipticity dispersion derived from the whole galaxy catalogue \citep{Giblin2021}. $\sigma_{\rm e}$ is the error of the ellipticity measurement defined in \citet{Hoekstra2002}. To suppress the dilution effect from the photo-$z$ uncertainties of the source galaxies, we remove the lens-source pairs with $z_{\rm s}-z_{\rm l}< 0.1$.
	
	We apply the correction of multiplicative bias to the measured excess surface density as
	\begin{equation}
		\Delta \Sigma^{\rm cal}(R)=\frac{\Delta \Sigma (R)}{1+K(z_{\rm l})} B(R)\,,
	\end{equation}
	where
	\begin{equation}
		1+K(z_{\rm l})=\frac{\sum_{\rm ls}\omega_{\rm ls}(1+m)}{\sum_{\rm ls}\omega_{\rm ls}}.
	\end{equation}
	where $m$ is the multiplicative bias as described in Sec.~\ref{ssec:DECaLS_survey}. The lensing signal is multiplied by boost factor $B(R) =n(R)/n_{\rm rand}(R)$, which is the ratio of the number density of sources relative to the number around random points, in order to account for dilution by sources that are  physically associated with lenses, and therefore not lensed \citep{Mandelbaum2005boostfactor, Mandelbaum2006_372_758}. Throughout this work, we use the Super W Of Theta (SWOT) code\footnote{\url{http://jeancoupon.com/swot}}~\citep{Coupon_2012} to calculate the excess surface density.
	
	\subsection{Joint constraint}
	The $\chi^2$ of lensing can be written as 
	\begin{equation}
		\chi^2_{\rm lensing} = \left[ \frac{\Delta\Sigma^{\rm ob}(R_i) - \Delta\Sigma^{\rm model}(R_i)}{\sigma_{\Delta\Sigma}} \right]^2
	\end{equation}

	We model the stacked excess surface density $\Delta\Sigma(R)$ as
	\begin{equation}
		\Delta \Sigma(R)=\Delta\Sigma_{\rm star}(R)+\Delta \Sigma_{\rm NFW}(R)+\Delta\Sigma_{\rm 2h}(R),
		\label{equ:model}
	\end{equation}
	where the first term represents the contribution of the central baryonic component, the second term represents the contribution of host projected dark matter haloes of the galaxies, and the third term represents the contribution of neighboring halos, namely the 2-halo terms.

	In many previous lensing analyses, the contribution of the baryonic component is often modeled as a point mass, the value of which is set to the sum of the stellar mass of the galaxies. However, the approximation is not accurate within around $2R_e$. Following \citetalias{Zhu_2023_paperI}, we assume the mass distribution of the baryonic component follows the r-band light distribution, the form of which is derived by fitting the Multi-Gaussian Expansion \citep[MGE;][]{Emsellem1994, Cappellari2002} to the r-band image of the galaxies in SDSS DR17. The excess surface density of the stellar component can be written as 
	\begin{equation}
		\Delta\Sigma_* (R)= \alpha_{\rm nsa}\Delta\Sigma_{\rm *,nsa}(R),
	\end{equation}
	where 
	\begin{equation}
		\Delta\Sigma_{\rm *,nsa}=\overline{\Sigma}_{\rm *,nsa}(<R)-\Sigma_{\rm *,nsa}(R),
	\end{equation}
    where $\Sigma_{\rm *,nsa}$ is the stellar mass distribution which follows the r-band MGE brightness distribution derived from \citetalias{Zhu_2023_paperI}, with normalization fixed to the stellar mass $M_{\rm *,nsa}$, which adopts a Chaberier IMF. $\alpha_{\rm nsa}$ is a normalization parameter which describes the probable mismatch between 
     $\Sigma_{\rm *,nsa}$ and the ground truth. In our "ggl only  model" and "ggl+dyn" model (defined below), we fix $\alpha_{\rm nsa}=1$.

	In our fiducial model, we use the NFW profile \citep{Navarro1996} to model the dark halo component:
	\begin{equation}
		\rho_{\rm NFW}(r) \propto \frac{1}{(r/r_{\rm s})(1+r/r_{\rm s})^2},
	\end{equation}
	where $r_{\rm s}$ is the scale radius where the local logarithmic slope $\frac{{\rm d}~\ln\rho}{{\rm d}~\ln r}=-2$, which can be derived from dark matter halo radius through the concentration  parameter $C_{\rm 200}=R_{\rm 200}/r_{\rm s}$. 
	
	The excess surface density $\Delta\Sigma_{\rm NFW}(R)$ can be calculated by integrating the three-dimensional density profile along the line of sight, which we assume aligned with the $z$ axis, as follow:
	
	\begin{equation}
		\begin{cases}
			\begin{aligned}
				\Sigma_{\rm NFW}(R)=\int_{-\infty}^{\infty} \rho_{\rm NFW}\left(\sqrt{R^2+z^2}\right)dz,\\\
				\overline{\Sigma}_{\rm NFW}(<R)=\frac{2}{R^2}\int_{0}^{R}R'\Sigma_{\rm NFW}(R')dR',\\
				\Delta\Sigma_{\rm NFW}(R)=\overline{\Sigma}_{\rm NFW}(<R)-\Sigma_{\rm NFW}(R).
			\end{aligned}
		\end{cases}
	\end{equation}
	For the NFW model, we use the analytical expression for $\Delta\Sigma$ presented in \cite{Wright1999astro.ph..8213O} to perform the model fitting.
	  The 2-halo term $\Delta\Sigma_{\rm 2h}$ can be written as
	\begin{equation}
		\Delta\Sigma_{\rm 2h} = \gamma_{\rm 2h} \ \Delta\Sigma_{\rm crit},
	\end{equation}
	where the tangential shear profile due to the neighboring halos \citep{Oguri_and_Hamana_2011} is:
	\begin{equation}
		\gamma_{\rm 2h}(\theta;M,z)=\int\frac{ldl}{2\pi}J_{\rm 2}(l\theta)\frac{\overline{\rho}_{\rm m}(z)b_{h}(M)}{(1+z)^3\Sigma_{\rm crit}D_{\rm A}^2(z)}P_{m}(k_{l};z),
	\end{equation}
	where $J_{\rm 2}$ is the second-order Bessel function, $\overline{\rho}_{\rm m}(z)$ is the mass density at $z$, $D_{\rm A}(z)$ is the angular diameter distance, $P_{\rm m}(k)$ is the linear matter power spectrum, $b_{\rm h}(M)$ is the halo bias derived by~\cite{Tinker2010}.
	
	As mentioned in  \autoref{subsec:dynamicalmodel}, we do not stack kinematic data as we do for the lensing data, instead, we incorporate the dynamical data of MaNGA galaxies by adding a term of $\chi^2_{\rm dyn}$ as
	\begin{equation}
		\chi^2_{\rm dyn} = \left( \frac{\langle M_{\rm dyn}(<R_0) \rangle- M_{\rm model}(<R_0)}{\sigma_{\rm M,dyn}}\right)^2 \,,
	\end{equation}
	
	Therefore, the total $\chi^2$ of the joint analysis can be written as
	\begin{equation}
		\chi^2_{\rm tot} = \chi^2_{\rm lensing} + \chi^2_{\rm dyn} \,.
	\end{equation}
    The averaged stellar mass within $R_{0}$ is given by
    \begin{equation}
    M_{*}(<R_{0})=\alpha_{\rm nsa}\frac{\left<M_{\rm *,nsa}\right>}{\left<M_{\rm *,JAM}\right>}M_{\rm *,JAM}(<R_{0}),
    \end{equation}
    where $M_{\rm *,JAM}(<R_{0})=\left<M/L\right>_{\rm *,JAM}L(<R_{0})$ is the stellar mass of JAM model within $R_{0}$. $\left<M/L\right>_{\rm *,JAM}$ is the stellar mass-to-light ratio ($M/L$) of JAM model and $L{(<R_{0}})$ is the luminosity within $R_{0}$ derived from Paper I.

	Different data sets and mass models are described as follows.
	\begin{itemize}
		\item{ggl only: only the weak gravitational lensing is used in fitting.  The stellar mass normalization $\alpha_{\rm nsa}=1$, and the model has two free parameters, $M_{\rm 200}$ and $C_{\rm 200}$.}
		\item{ggl+dyn: both dynamical data and weak gravitational lensing signal are used in fitting. The stellar mass normalization $\alpha_{\rm nsa}=1$.}
		\item{ggl+dyn+free ml: both dynamical data and weak gravitational lensing signal are used in fitting. The mass model has three free parameters, $M_{\rm 200}$, $C_{\rm 200}$, and the stellar mass normalization $\alpha_{\rm nsa}$. }
		\item{ggl+dyn+gnfw: similar to ggl+dyn, but the gNFW model describe is used instead of NFW model for the host halo component. The stellar component is fixed as $\alpha_{\rm nsa}=1$. }
	\end{itemize}
	
	We adopt the Markov chain Monte Carlo (MCMC) technique to perform the modeling fitting and calculate the posterior distribution. We use 24 chains of 300,000 steps with the MCMC ensemble sampler \textsc{Emcee}\footnote{\url{https://emcee.readthedocs.io/en/stable/}} \citep{2013PASP..125..306F}.

    Studying galaxy properties in sub-sample bins with stacked galaxy-galaxy gravitational lensing method may introduce biases \citep{Sonnenfeld_and_Leauthaud_2018}. We performed a test using halo masses that are identical to the sub-samples used in this work and assumed they follow an NFW profile distribution and a mass-concentration relation. We then stacked the signals and fit them using the NFW model. The results showed a bias between the best-fit model parameters ($M_{\rm 200}$, $C_{\rm 200}$) and the true mean value. However, the true mean values were consistent with the fitting results within the $1\sigma$ error range. Therefore, the bias introduced by the stacked method will not change our conclusion.
	
	%%%%%%%%%%%%%%%%%%%%%%%%%%%%%%%%%%%%%%%%%%%%%%%%%%
	\section{Results}
	\label{sec:result}
\begin{figure*}
		\centering
		\includegraphics[width=0.8\textwidth]{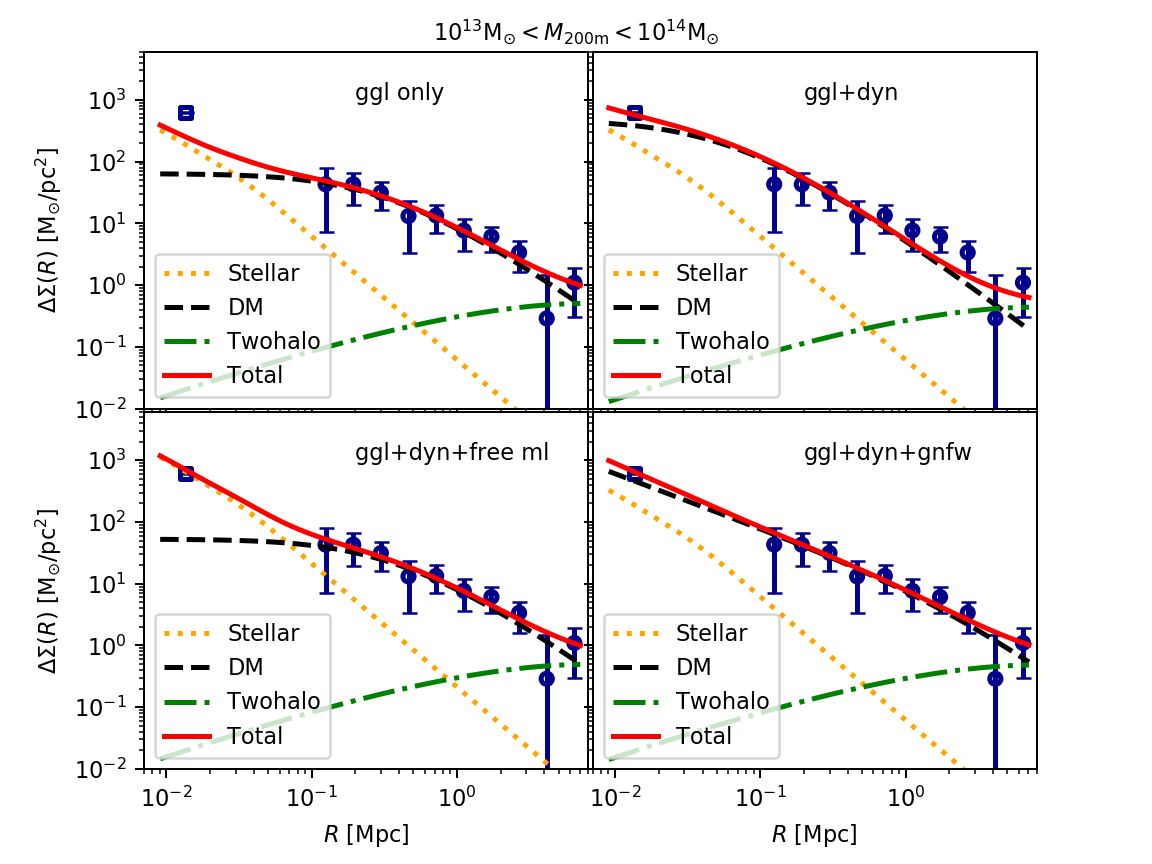}
		\caption{The figure shows the observational lensing data for group mass bin of  $10^{13}{\rm M_{\odot}} <M_{\rm 200m}<10^{14}{\rm M_{\odot}}$. The dark blue circles with error bars show the measured $\Delta\Sigma(R)$, and the blue squares with error bars show the predicted $\Delta\Sigma(R)$ by the mass model derived from MaNGA dynamical modeling alone. This 2D dynamical lensing signal is not involved in the model fitting and we just show it here for illustration. We show the best fit for different mass models in different panels, where the labels of the mass models are marked.  The solid lines, dotted lines, dashed lines, and dashdot lines show the total signal, and the contribution of the stellar component, the host dark matter halo, and the two-halo term, respectively.}
		\label{fig:Bestfit_13_14}
	\end{figure*}

\begin{figure*}
		\centering
		\includegraphics[width=0.8\textwidth]{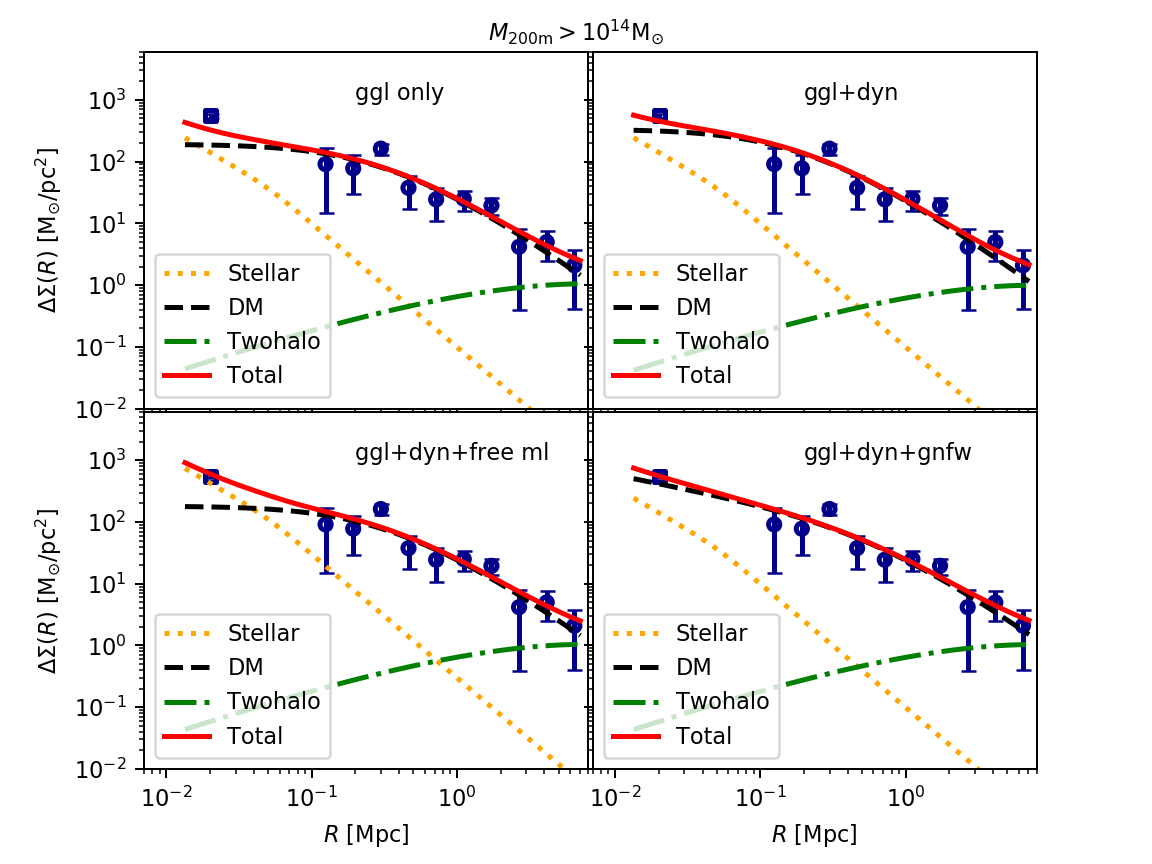}
		\caption{Similar to Fig.~\ref{fig:Bestfit_13_14}, but for the cluster mass bin of $M_{\rm 200m}>10^{14}{\rm M_{\odot}}$.}
		\label{fig:Bestfit_14_16}
	\end{figure*}

\begin{figure*}
		\centering
		\includegraphics[width=\columnwidth]{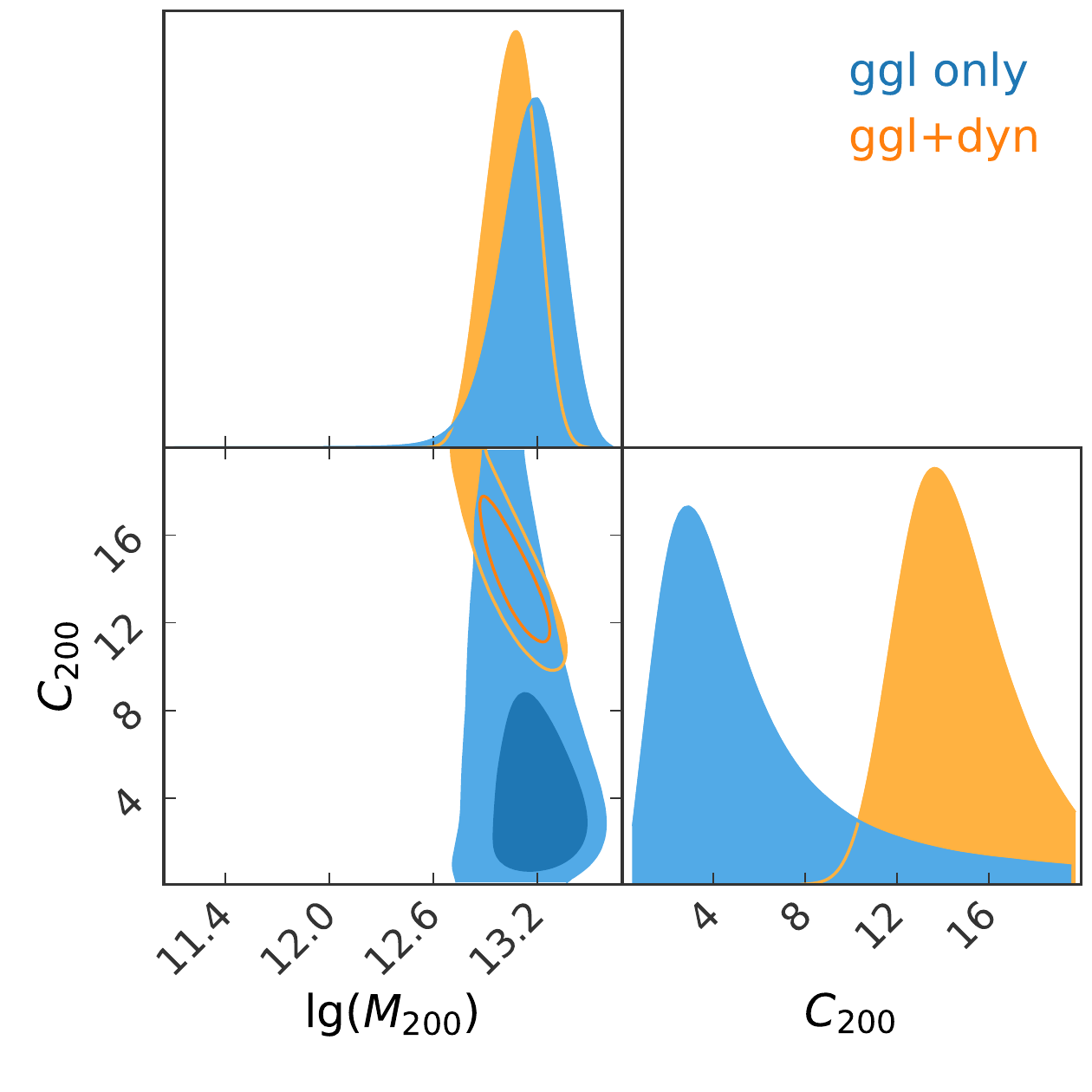}
		\includegraphics[width=\columnwidth]{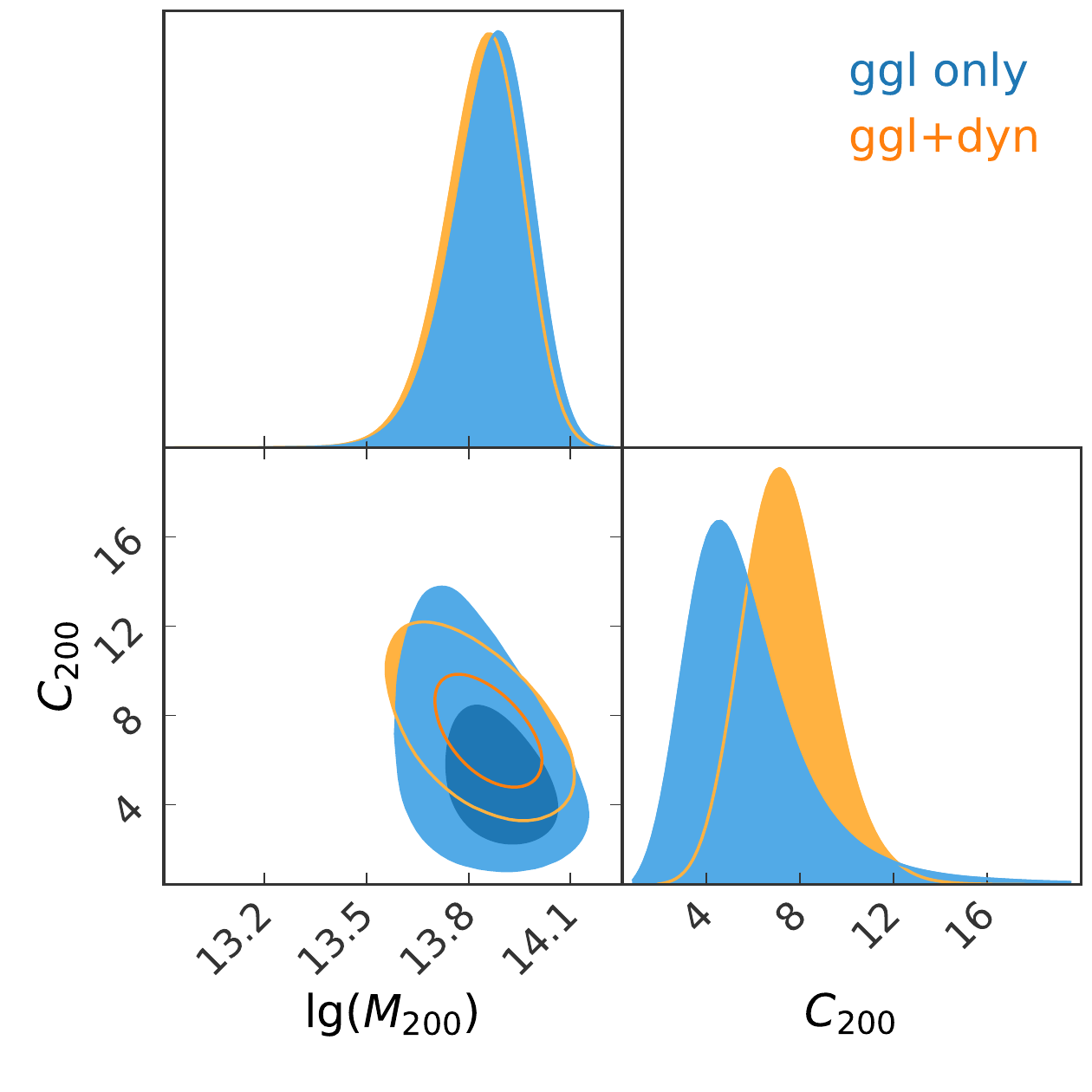}\\
		\caption{The figure shows the posterior distribution for ggl only(blue) and ggl+dyn (orange) cases. The left panel shows the results of the $10^{13}{\rm M_{\odot}} <M_{\rm 200m}<10^{14}{\rm M_{\odot}}$ bin, and the right panel shows that of the $M_{\rm 200m}>10^{14}{\rm M_{\odot}}$ bin. The contours show  68\% and 95\% confidential levels. $M_{\rm 200}$ is in units of $\rm {M_{\odot}}$.}
		\label{fig:posteriors_corner}
	\end{figure*}

\begin{figure*}
		\centering
		\includegraphics[width=\columnwidth]{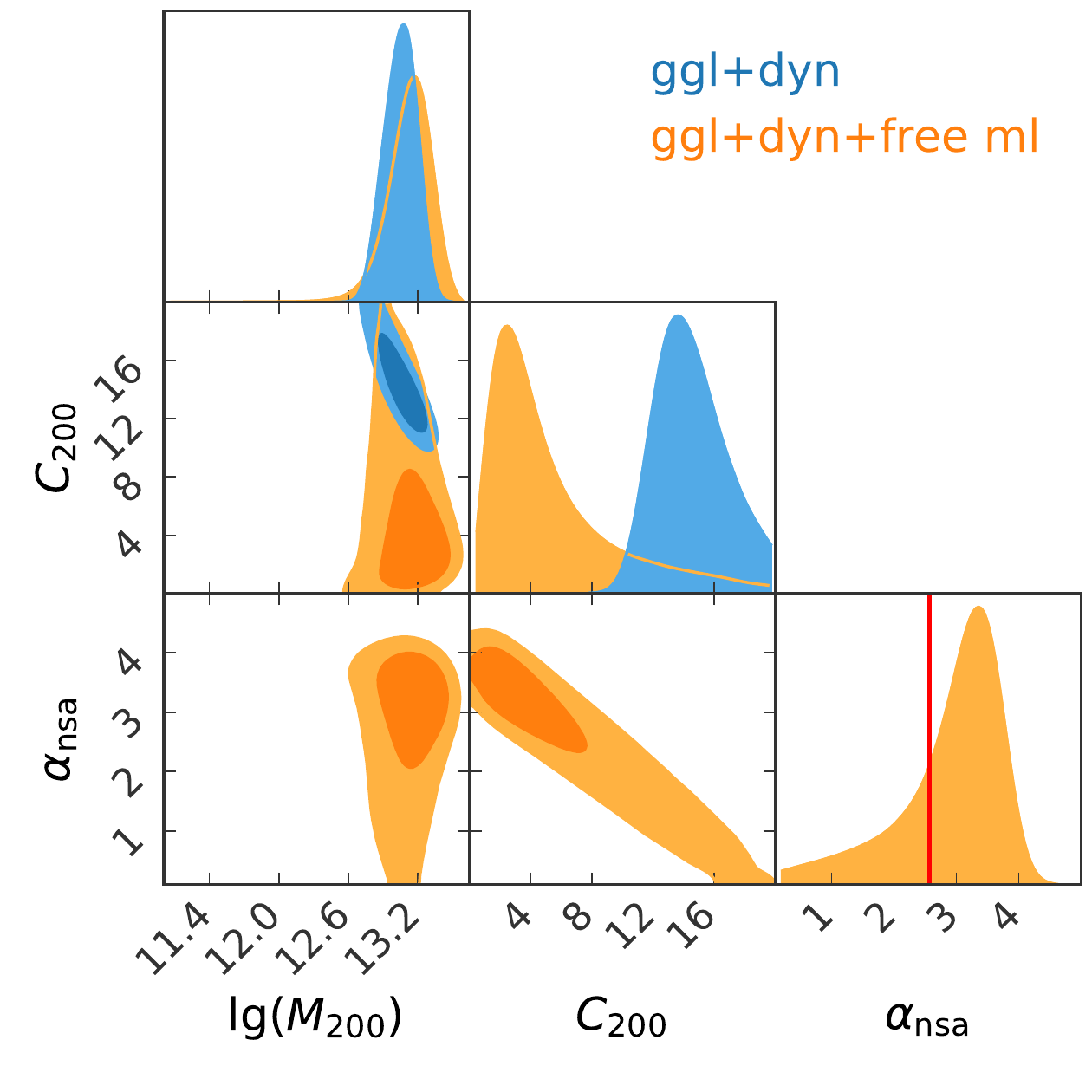}
		\includegraphics[width=\columnwidth]{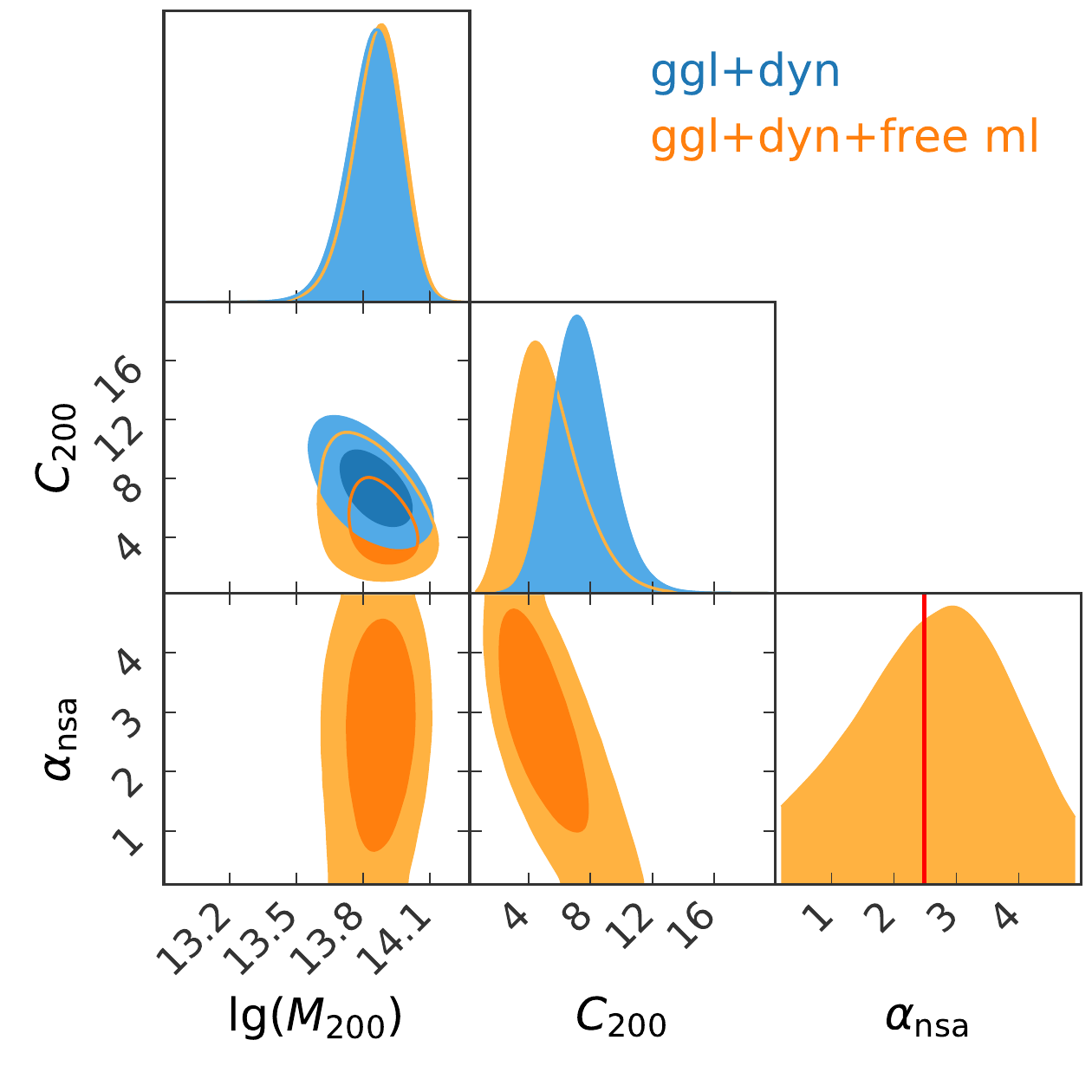}\\
		\caption{Similar to Fig.~\ref{fig:posteriors_corner}, but comparing the ggl+dyn model and the ggl+dyn+free ml model. The vertical solid lines represent the mean value of $\frac{M_{\rm *, JAM}}{M_{\rm *, nsa}}$ of MaNGA galaxies derived by \citetalias{Zhu_2023_paperI}, which uses kinematic data alone.}
		\label{fig:posteriors_corner_ml}
	\end{figure*}

\begin{figure*}
		\centering
		\includegraphics[width=\columnwidth]{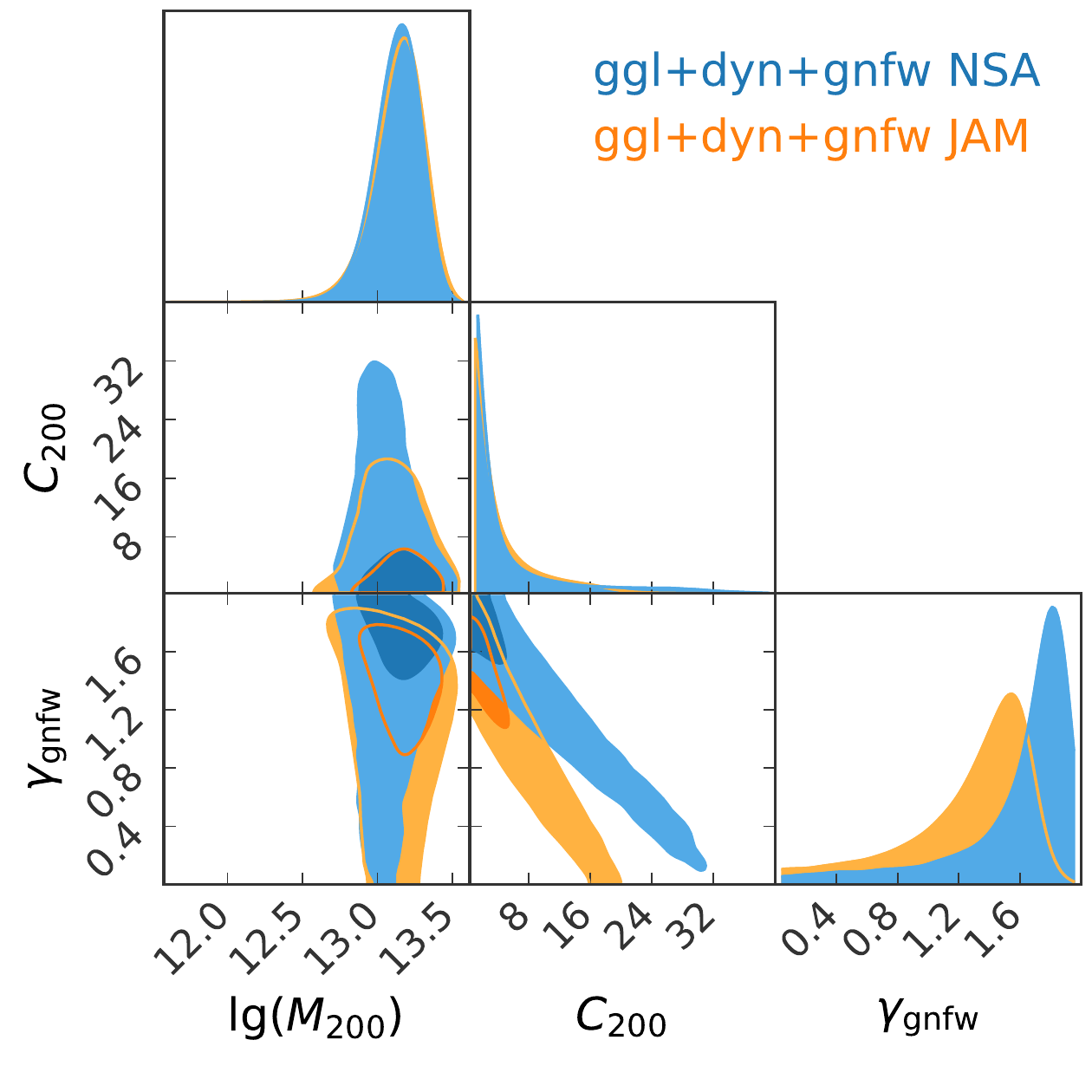}
		\includegraphics[width=\columnwidth]{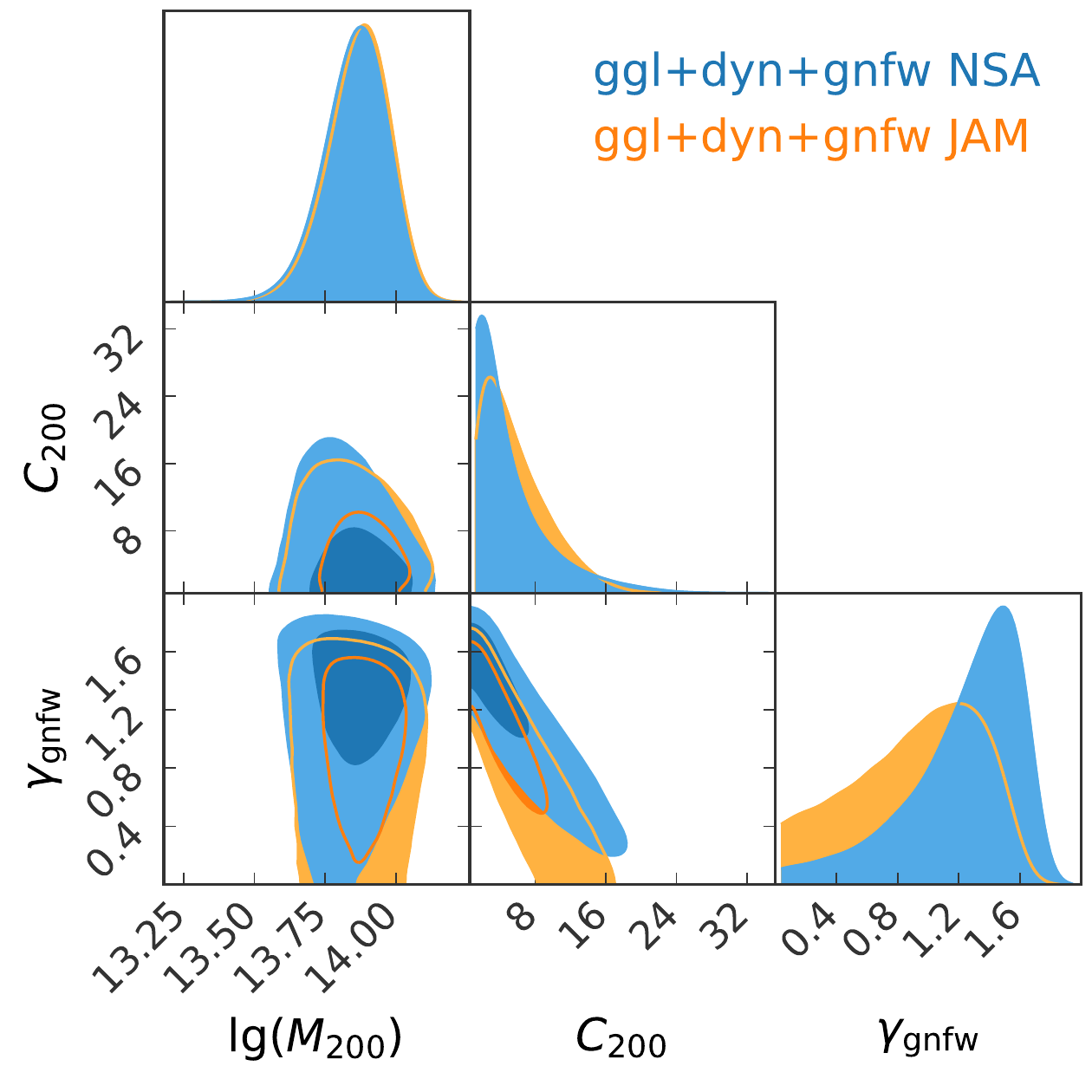}\\
		\caption{The figure shows the posterior distribution of ggl+dyn+gnfw model with different stellar mass model (blue: NSA; orange:JAM). The left panel shows the results of the $10^{13}{\rm M_{\odot}} <M_{\rm 200m}<10^{14}{\rm M_{\odot}}$ bin, and the right panel shows that of the $M_{\rm 200m}>10^{14}{\rm M_{\odot}}$ bin. The contours show  68\% and 95\% confidential levels. $M_{\rm 200}$ is in units of $\rm {M_{\odot}}$.}
		\label{fig:posteriors_corner_gnfw}
	\end{figure*}

	We present the gravitational lensing signal and the best-fit models in Figs.~\ref{fig:Bestfit_13_14} and \ref{fig:Bestfit_14_16} for the two halo mass bins. We also show the decomposed components of the best-fit model, namely the dark matter halo (dashed), stellar component (dotted), and the two halo term (dashdot). Each figure contains four different panels, showing the fitting results of four different models. The posterior distribution of the model parameters can be found in Figs.~\ref{fig:posteriors_corner}, \ref{fig:posteriors_corner_ml}, and \ref{fig:posteriors_corner_gnfw}. The best-fit parameters with uncertainties are listed in Table~\ref{tab:fitting_result} and Table~\ref{tab:fitting_result_gnfw}.
	
	We derive mean halo mass of ${\rm lg}(M_{\rm 200}{\rm [M_{\odot}]})= {13.07}^{+0.13}_{-0.14}$ and ${13.87}^{+0.09}_{-0.12}$ for the group and the cluster bins respectively (ggl+dyn model). The best-fit values of the halo  mass are stable among different mass model assumptions, with differences within 1$\sigma$ error, which reflects the fact that the total mass of a halo is mainly constrained by the weak lensing data which measures the density distribution at a larger scale.

	%  fig2    

	%  fig3       

	%fig4

	%fig5

	%fig

\begin{figure*}
		\begin{center}
			\includegraphics[width=2\columnwidth]{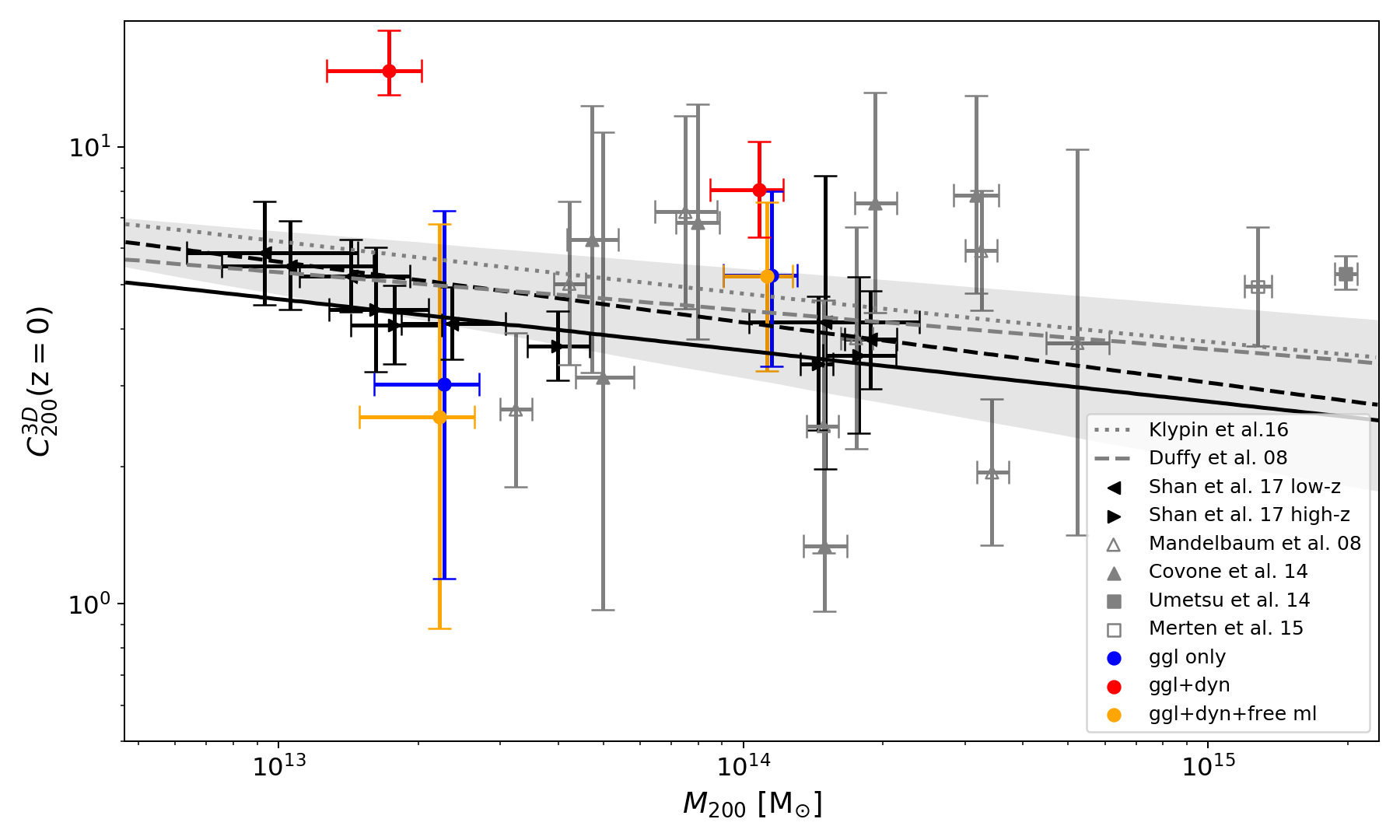}
			
			\caption{The comparison between our best-fit parameter result and the mass-concentration relation in \protect\cite{Shan2017}. All the concentrations are corrected from 2D to 3D and rescaled to $z = 0$, assuming the redshift evolution from \protect\cite{Klypin2016}. The blue, red, and orange circles with error bars show the fitting result of `ggl only', `ggl+dyn', and `ggl+dyn+free ml' models. In \protect\cite{Shan2017}, they binned the lens samples into two redshift bins, low-$z$ ($0.2<z<0.4$) and high-$z$ ($0.4<z<0.6$) shown with black triangles left and black triangles right. The black solid and dashed line shows the best-fit mass-concentration relation for the low-$z$ and high-$z$ subsamples. The grey shaded area shows the 1$\sigma$ uncertainty of this relation of high-$z$ subsamples. The grey dashed and dotted curves are the simulation predictions by \protect\cite{Duffy2008} and \protect\cite{Klypin2016}. The grey symbols denote the lensing-based measurements of concentration and mass by \protect\cite{Mandelbaum2008} with SDSS (median redshift $(\overline{z} \sim 0.22)$; \protect\cite{Covone2014} with CFHTLenS $(\overline{z} \sim 0.36)$; and \protect\cite{Umetsu2014} $(\overline{z} \sim 0.35)$ and \protect\cite{Merten2015} $(\overline{z} \sim 0.40)$ with the CLASH cluster sample. The data points of \protect\cite{Merten2015} were binned.}
			\label{fig:m-c-relation}
		\end{center}
	\end{figure*}

	On the other hand, the total density profiles at the inner~100 kpc region depend strongly on whether stellar kinematics is included in the fitting. If we use weak lensing alone to constrain the mass
	model with an NFW profile and the stellar mass normalization fixed with NSA catalogue,   we get $C_{\rm 200}=2.73^{+3.80}_{-1.70}$ for $10^{13}-10^{14}$ $M_{\odot}$ bin, and $C_{\rm 200}=4.54^{+2.39}_{-1.66}$ for $>10^{14}M_{\odot}$ bin, while the best-fit model predicts an amplitude of $\Delta \Sigma$ significantly lower than the result from MaNGA stellar kinematics. If we use the same mass model to match both inner stellar kinematic data and the weak lensing data (ggl+dyn case), the values of $C_{\rm 200}$ raise to $13.26_{-1.52}^{+3.02}$, and $7.01_{-1.49}^{+1.90}$ for the two bins respectively,
	significantly higher than that predicted by N-body cosmological simulations \citep{Duffy2008} as well as other weak lensing measurements (see Fig.~\ref{fig:m-c-relation}).

	The discrepancy between the weak lensing and stellar kinematics can be alleviated by allowing the variation of stellar mass contribution, by relaxing the assumption of a universal IMF, or the inner density slope of dark matter halo. In the bottom panels of Figs.~\ref{fig:Bestfit_13_14} and \ref{fig:Bestfit_14_16}, we show the fitting results of these two new mass models. One can find that with additional free parameters, the mass model can fit both data set well.

	In Fig.~\ref{fig:posteriors_corner_ml}, we show the posterior distribution for the mass model with a free stellar mass normalization parameter. By combining the weak lensing and stellar kinematic data, we find best-fit values $\alpha_{\rm nsa}=3.34_{-0.73}^{+0.52}$ for the group bin and $\alpha_{\rm nsa}=2.87_{-1.38}^{+1.12}$ for the cluster bin. In both cases, an $M/L$ $\sim$ 3 times higher than that given by the NSA catalogue is preferred. In the figure, we also mark the mean stellar normalization derived from the \citetalias{Zhu_2023_paperI} catalogue using the red vertical line, where the decomposed stellar mass distribution is derived from stellar kinematics alone. For the group bin, the best-fit value of stellar mass normalization of this work is slightly higher than that derived from \citetalias{Zhu_2023_paperI}, but consistent at $\sim 1\sigma$ level. For the cluster bin, the two results agree well with each other.
	
	In this project, we make the assumption that the stellar mass surface density is proportional to the MGE surface brightness model and normalize it to the stellar mass derived from the NSA catalogue. However, it should be noted that the total luminosity used to derive the stellar mass by the NSA catalogue and that from the MGE model may differ slightly, which can also contribute to the normalization factor $\alpha_{\rm nsa}$ value but this difference does not affect the results of "ggl+dyn+ml free" model, thus won't change the conclusion of this paper. 
 
	Allowing the variation of stellar mass normalization has a direct impact on the best-fit value of the concentration parameter for the group bin. When using a free $\alpha_{\rm nsa}$, the concentration parameter decreases from $C_{\rm 200}=13.26_{-1.52}^{+3.02}$  to $C_{\rm 200}=2.32^{+3.78}_{-1.52}$ which is close to the value from using galaxy-galaxy lensing alone. The best-fit value of concentration for the cluster mass bin also decreases, but not significantly. 
	
	In Fig.~\ref{fig:m-c-relation}, we compare concentration derived from this work with mass-concentration relations measured in weak lensing surveys and numerical simulations \citep{Shan2017, Mandelbaum2008, Duffy2008, Klypin2016, Umetsu2014, Covone2014, Merten2015}.
	Here, we correct the 2D concentration to 3D concentration with the relation $C_{\rm 2D}(M)=C_{\rm 3D}(M)\times 1.630M^{-0.018}$ provided by \cite{Giocoli2012}, who found that halo triaxiality and substructures within the host halo virial radius can bias the observed 2D mass–concentration relation. The concentrations are all scaled to $z=0$ by dividing $(1+z)^{-0.67}$, assuming the redshift evolution
	from \cite{Klypin2016}. One can find that the concentration derived with stellar mass normalization fixed to the NSA catalogue ("ggl+dyn" ) is significantly higher than the previous lensing observation and the prediction of numerical simulation, while the results derived with "ggl+dyn+free ml" agree with these previous measurements within 1$\sigma$. 

 Different models also predict different dark matter fraction for the central region. Table~\ref{tab:fitting_result} lists the dark matter fraction within the average $R_{\rm e}$ of the sample for different models as $f_{\rm dm}(<\langle R_{\rm e} \rangle)$.  For both mass bins, the "ggl+dyn" model yields a high dark matter fraction, with $f_{\rm dm}(<\left \langle R_{\rm e} \right\rangle)=0.66^{+0.02}_{-0.03}$ for the group bin and $f_{\rm dm}(<\left \langle R_{\rm e} \right\rangle)=0.68^{+0.05}_{-0.05}$ for the cluster bin. In contrast, the "ggl+dyn+free ml" model predicts much lower values, with $f_{\rm dm}(<\left \langle R_{\rm e} \right\rangle)=0.06^{+0.14}_{-0.04}$ for the group bin and $f_{\rm dm}(<\left \langle R_{\rm e} \right\rangle)=0.23^{+0.24}_{-0.11}$ for the cluster bin, which agrees with the values derived from stellar kinematic data alone in \citetalias{Zhu_2023_paperI}.

	%  Fig6
	
	In Figs.~\ref{fig:Bestfit_13_14} and \ref{fig:Bestfit_14_16}, we also investigate whether a steeper asymptotic inner density slope can also explain the observational data. Since the inner density slope degenerates strongly with stellar mass normalization parameter, we choose to set the latter to two fixed values during the fitting, the NSA value, where $\alpha_{\rm nsa}=1$; the value derived with JAM model from \citetalias{Zhu_2023_paperI}, where $\alpha_{\rm nsa}=2.57$ and $2.49$ for the group and the cluster bins respectively. We present the best-fit asymptotic density slope, $\gamma_{\rm gnfw}$ in Table \ref{tab:fitting_result_gnfw}. If we fix the stellar mass normalization to that derived from the NSA catalogue, we obtain a best-fit value of density slope $\gamma_{\rm gnfw}=1.82_{-0.25}^{+0.15}$ and $\gamma_{\rm gnfw}= 1.48_{-0.41}^{+0.2}$ for the group and the cluster bins respectively. The case of the NFW profile ($\gamma = 1$) is at the 10th percentile and 23th percentile of the posterior distribution of $\gamma_{\rm gnfw}$ for the group and cluster bins. If we choose to fix the stellar mass normalization  to the mean value derived from JAM instead, we obtain $\gamma_{\rm gnfw}=1.57_{-0.43}^{+0.16}$ ($\gamma=1$ is the 21th percentile ), and $\gamma_{\rm gnfw}=1.21_{-0.59}^{+0.28}$ ($\gamma=1$ is 53th percentile).
 
\begin{figure*}
		\centering
		\includegraphics[width=0.85\columnwidth]{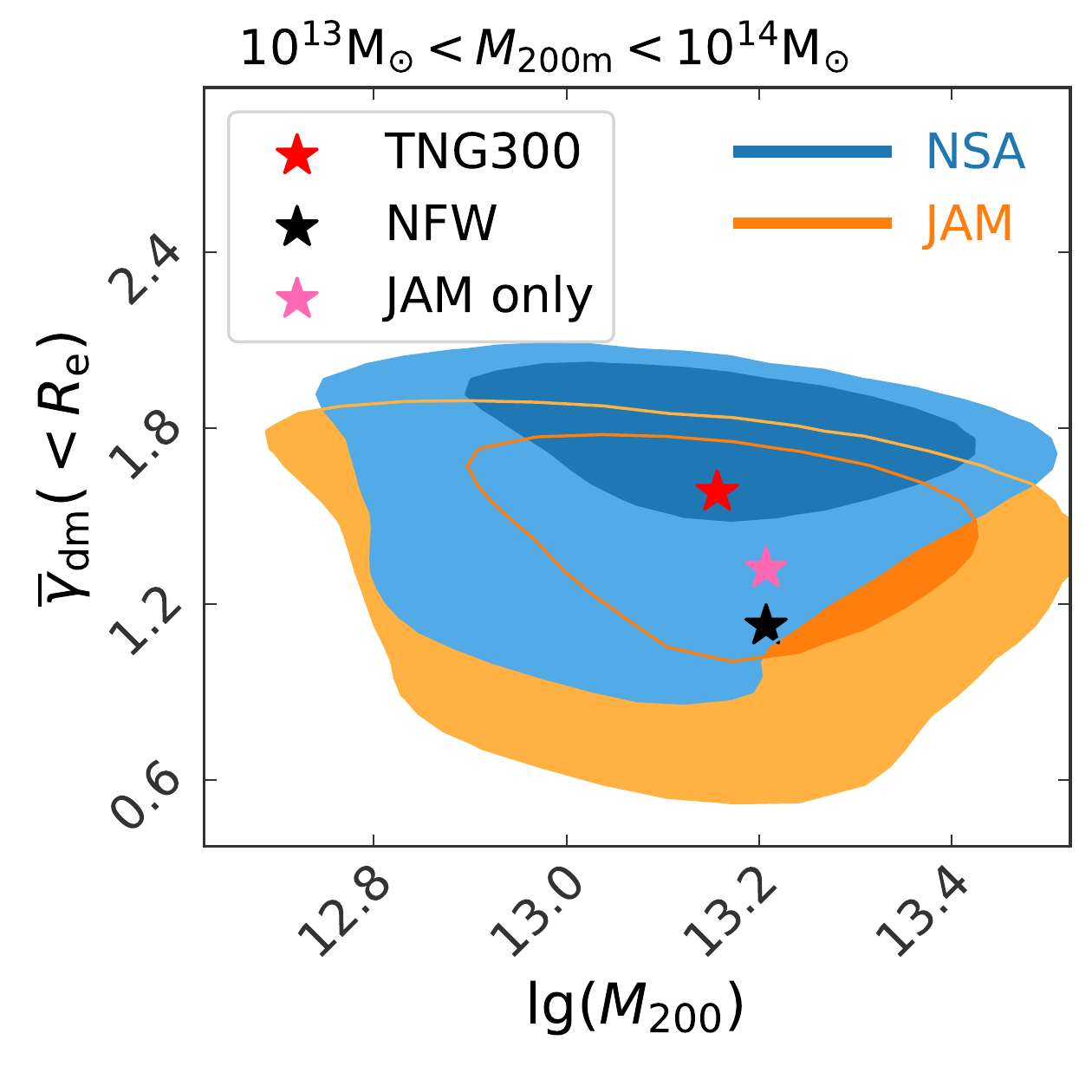}
		\includegraphics[width=0.85\columnwidth]{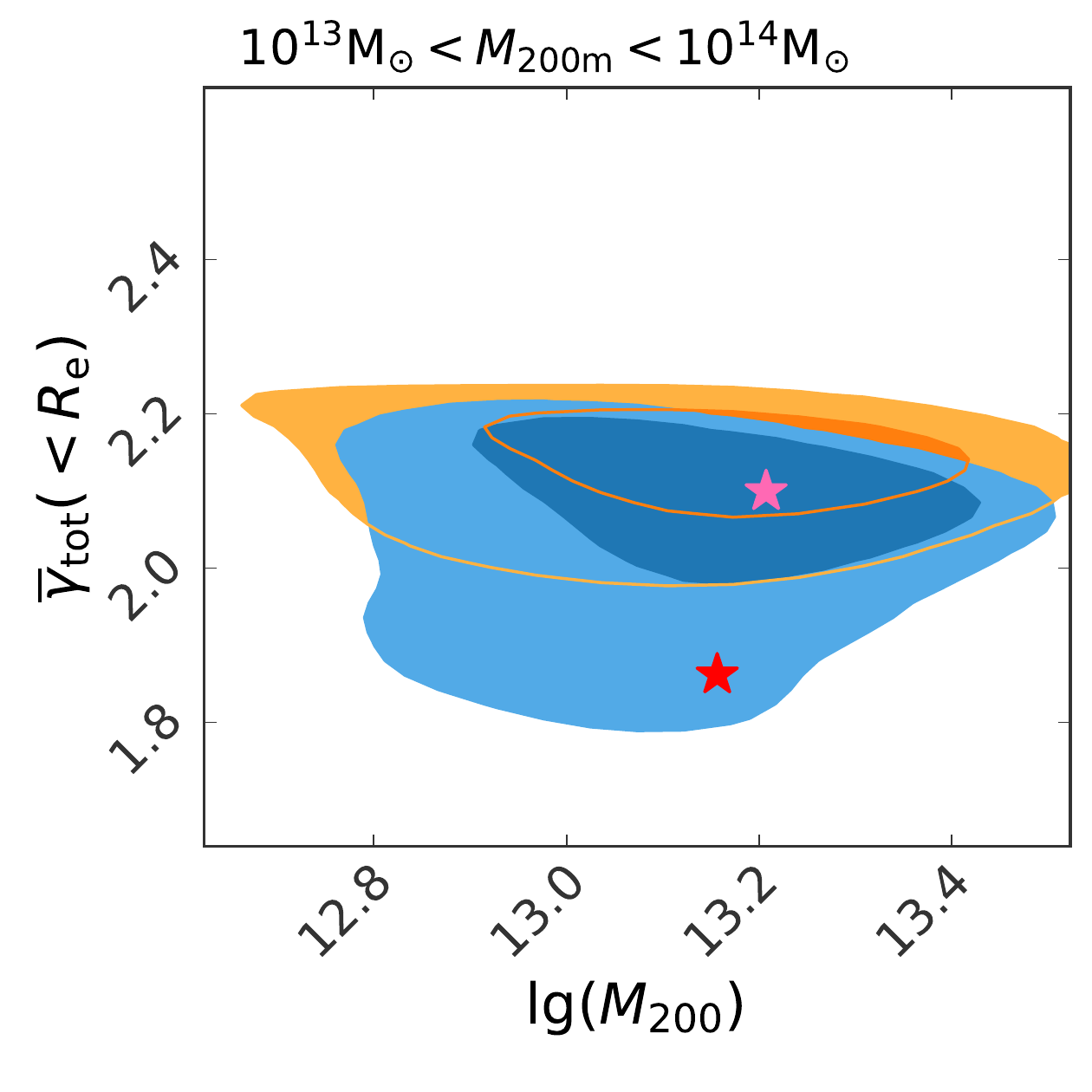}\\
		\includegraphics[width=0.85\columnwidth]{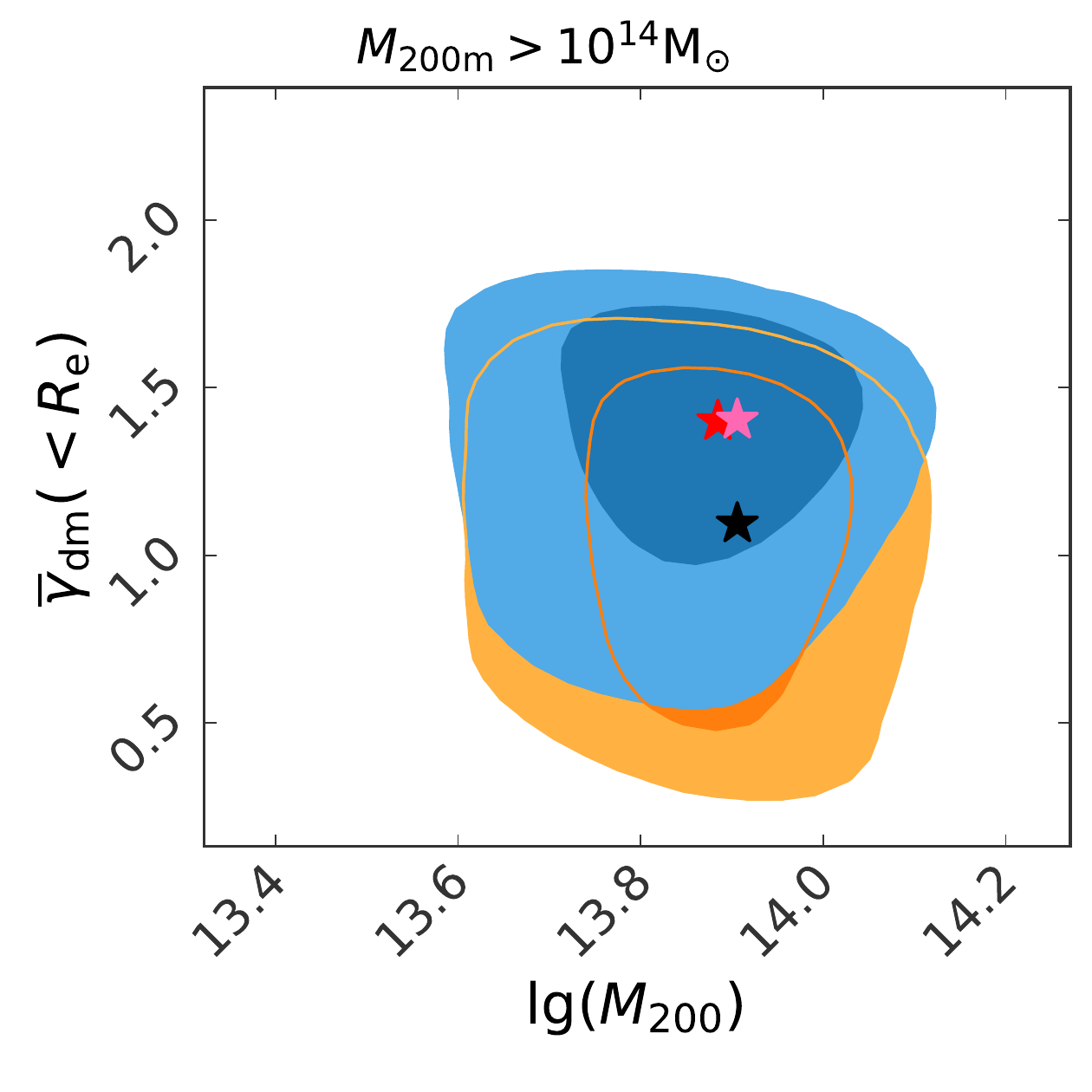}
		\includegraphics[width=0.85\columnwidth]{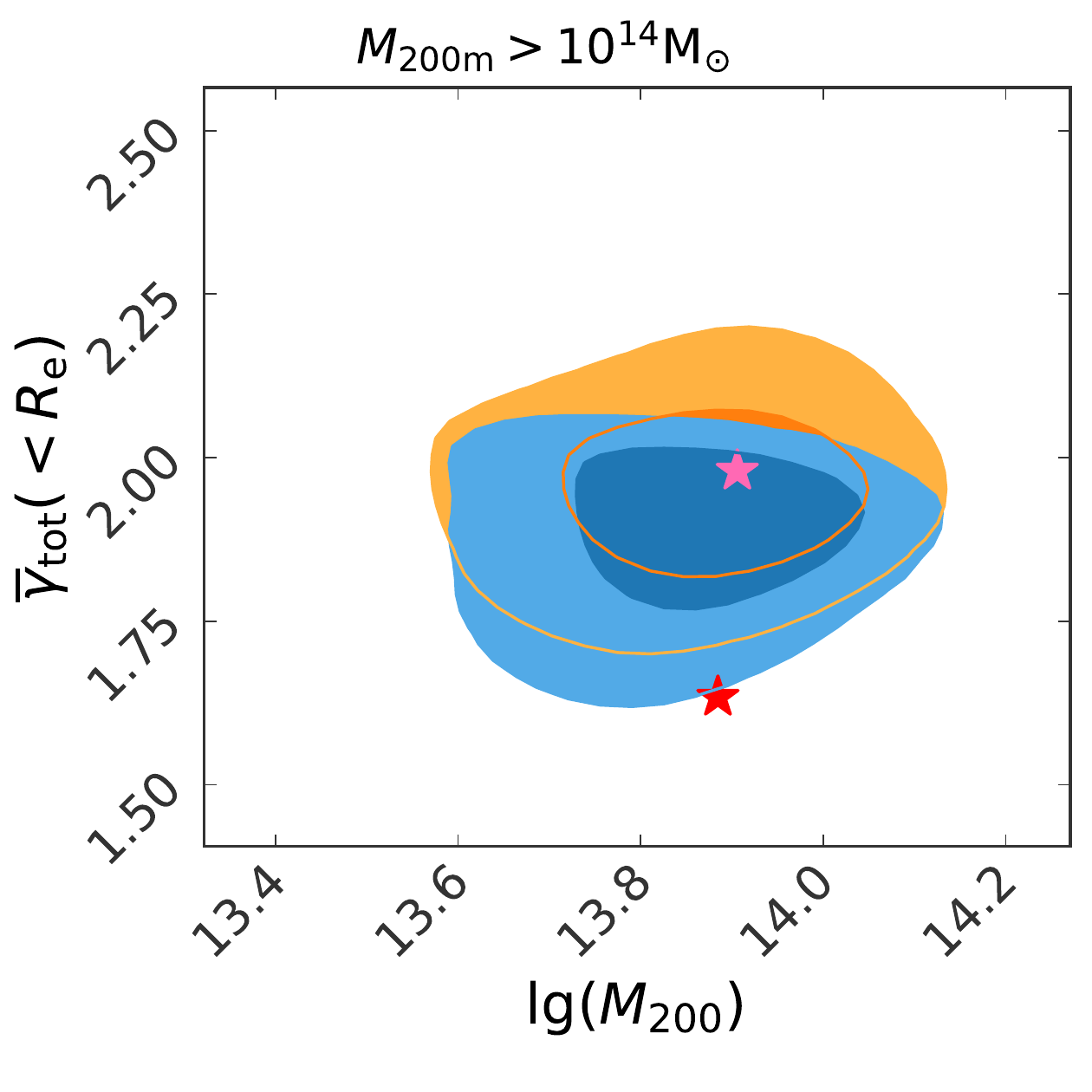}
		\caption{2D distribution of the inner density slope at the scale of $R_{\rm e}$ and dark matter halo mass $M_{\rm 200}$ (in units of $M_{\odot}$). The results of $10^{13}{\rm M_{\odot}} <M_{\rm 200m}<10^{14}{\rm M_{\odot}}$ and $M_{\rm 200m}>10^{14}{\rm M_{\odot}}$ subsample bins are shown in the upper and lower panels, respectively. The dark matter inner density slope and total mass inner density slope at the scale of $R_{\rm e}$ are shown in the left and right panels. In all panels, the blue lines show the fitting results using the NSA stellar mass in the model fitting and the orange lines show results using the JAM stellar mass. The red star in each panel is the corresponding mean value estimated from the TNG300 simulation. The black stars in the left panels are mass-weighted inner density slope of dark matter when we assume that an NFW profile has the best-fit dark halo mass of our sub-samples and follows the mass-concentration relation from \protect\cite{Duffy2008}. Effective radius of the NFW profile are set equal to the average effective radius of the observed sub-samples. Pink stars show the stacked mass-weighted inner density slope of JAM model from Paper I.}
		\label{fig:inner_density_slope}
	\end{figure*}

	For the gNFW profile, the inner density slope degenerates with scale radius $r_{\rm s}$ \citep{Dutton2014, He2020}, so we also calculate mass-weighted density slope within $R_{\rm e}$ to quantify the shape of the density profile, where
	\begin{equation}
		\overline{\gamma}(<R_{\rm e})=-\frac{1}{M(R_{\rm e})}\int_{0}^{R_{\rm e}}4\pi r^2\rho(r)\frac{\rm{dlog}\rho}{\rm{dlog}{\it r}}\rm{d}r=3-\frac{4\pi {\it R_{\rm e}}^{3}\rho({\it R_{\rm e} })}{{\it M}({\it R}_{\rm e})},
    \label{equ:mass_weighted_density_slope}
	\end{equation}
	where $\rho(r)$ is the mass density and the $M(R_{\rm e})$ is the mass within the radius $\it R_{\rm e}$. The best-fit values of $\overline{\gamma}$ are shown in Table \ref{tab:fitting_result_gnfw}, and in the left panels of Fig.~\ref{fig:inner_density_slope}, we show the posterior distribution of mass-weighted density slope of dark matter component. For the group bin, $\overline{\gamma}_{\rm dm}(<R_{\rm e})$ disfavor the NFW model prediction if we set $\alpha_{\rm nsa}=1$, but the two agree within $1\sigma$, if $\alpha_{\rm nsa}$ is set to the values from \citetalias{Zhu_2023_paperI}. For the cluster bin, $\overline{\gamma}_{\rm dm}(<R_{\rm e})$ is not tightly constrained, thus the predictions from models with different $\alpha_{\rm nsa}$ setting broadly agree with each other. 
	
	In Fig.~\ref{fig:inner_density_slope}, we also show the posterior distribution for the mass-weighted total density slope. As expected, the best-fit total density slope does not change with the choice of  $\alpha_{\rm nsa}$. For the lower mass bin, the total density slope $\overline{\gamma}_{\rm tot}(<R_{\rm e})$ is $2.12_{-0.09}^{+0.05}$, slightly steeper than the singular isothermal case, while the slope for the cluster mass bin is flatter ($1.93_{-0.10}^{+0.06}$), which may due to the increasing dark matter fraction in the cluster scale haloes. 

    In Fig.~\ref{fig:inner_density_slope}, we also present the mass-weighted density slopes obtained from the cosmological hydrodynamical simulation, specifically the IllustrisTNG project (referred to as TNG300), with red stars. We select central galaxies from the TNG300 simulation (snapshot=99, $z=0$) whose FoF halo mass  $M_{\rm 200}$ falls within the $1\sigma$ range of the halo masses in our lens sub-samples. The effective radius of the central galaxies in the $r$ band, projected along the $z$-axis, is obtained from  \citet{Genel2018MNRAS.474.3976G}. The enclosed mass $M(R_{\rm e})$ and the mass density $\rho(R_{\rm e})$ in Equ.~\ref{equ:mass_weighted_density_slope} are directly calculated from the simulation data without any mass density model fitting. For both cluster and group mass bins, the mass-weighted average dark matter density slopes from TNG300 are consistent within the 1$\sigma$ confidence levels with those predicted by mass models with differently fixed $\alpha_{\rm nsa}$. The mass-weighted total density slope is consistent with prediction of TNG300 at 2$\sigma$ confidence levels when $\alpha_{\rm nsa}=1$, while the prediction of TNG300 is $2\sigma$ lower than that of our model fitting when $\alpha_{\rm nsa}$ is set to the values from Paper I. 
    
    In Fig.~\ref{fig:inner_density_slope}, the pink stars represent the stacked mass-weighted density slopes of the solely stellar kinematic data with JAM model fitting in Paper I. Results obtained solely from stellar kinematic fitting are consistent within a $1\sigma$ range with those when $\alpha_{\rm nsa}$ is set to the values from Paper I.

	\begin{table*}
\begin{center}
\caption{Posterior constraints of ggl+dyn+gnfw models. The first three columns, show the halo mass range, the number of lenses, and the stellar mass model. The following columns show the fitting parameters, halo mass, inner density slope, mass weighted dark matter density slope, mass weighted total density slope, and dark matter fraction within the mean value of $R_{\rm e}$.}
\begin{tabular}{llllllll}
 \hline  \hline
 Halo mass $M_{\rm 200m}$ range
 & Nlens
 & $M_{*}$ model
 &${\rm lg(}M_{\rm 200}{\rm [M_{\odot}])}$
 & $\gamma_{\rm gnfw}$
 & $\overline{\gamma}_{\rm dm}(<R_{\rm e})$
 & $\overline{\gamma}_{\rm tot}(<R_{\rm e})$
 & $f_{\rm dm}(<\left<R_{\rm e}\right>)$\\
 \hline
$10^{13}{\rm M_{\odot}} <M_{\rm 200m}<10^{14}{\rm M_{\odot}}$
& 422
& NSA
& $13.16_{-0.15}^{+0.13}$
& $1.82_{-0.25}^{+0.15}$
& $1.83_{-0.22}^{+0.13}$
& $2.12_{-0.09}^{+0.05}$
& $0.69_{-0.04}^{+0.02}$\\
$10^{13}{\rm M_{\odot}} <M_{\rm 200m}<10^{14}{\rm M_{\odot}}$
& 422
& JAM
& $13.2_{-0.18}^{+0.11}$
& $1.57_{-0.43}^{+0.16}$
& $1.57_{-0.38}^{+0.14}$
& $2.15_{-0.05}^{+0.04}$
& $0.26_{-0.06}^{+0.03}$\\
$M_{\rm 200m}>10^{14}{\rm M_{\odot}}$
& 97
& NSA
& $13.92_{-0.16}^{+0.05}$
& $1.48_{-0.41}^{+0.2}$
& $1.55_{-0.38}^{+0.13}$
& $1.93_{-0.1}^{+0.06}$
& $0.73_{-0.07}^{+0.04}$\\
$M_{\rm 200m}>10^{14}{\rm M_{\odot}}$
& 97
& JAM
& $13.92_{-0.14}^{+0.06}$
& $1.21_{-0.59}^{+0.28}$
& $1.14_{-0.41}^{+0.31}$
& $1.95_{-0.09}^{+0.08}$
& $0.36_{-0.11}^{+0.09}$\\

\hline
\hline
\end{tabular}
\label{tab:fitting_result_gnfw}
\end{center}
\end{table*}

	\section{Discussion and Conclusions}
	\label{sec:discussion_and_conclusion}
	
	In this paper, we presented the joint analysis of the stellar kinematic data and the galaxy-galaxy lensing data for a sample groups and clusters in the overlapping region between the MaNGA survey and the DECaLS DR8 imaging survey. Combining the two data sets allows us to derive the mean halo mass and measure the radial density profile from ~10 kpc to Mpc for the two sample bins. Intriguingly, we find the excess surface density derived using stellar kinematics cannot be naturally explained by adding an NFW halo derived from galaxy-galaxy lensing alone to a stellar mass component whose normalization is fixed to the value derived by the NSA catalogue.

	To match the high surface density derived from stellar kinematic data, the best-fit NFW profile requires concentration parameter $C_{\rm 200}=13.26_{-1.52}^{+3.02}$ (group bin) and $C_{\rm 200}=7.01_{-1.49}^{+1.9}$ (cluster bin), which is significantly higher than that predicted by the mass-concentration relation derived by \citet{Duffy2008} using cosmological numerical simulation.

	By setting the stellar mass normalization as a free parameter, we obtain a better fit to the observational data and draw the concentration parameter to the normal amplitude. We find that observational data allows us to put constraints on the stellar mass normalization $\alpha_{\rm nsa}$. In both the group and the cluster bins, the fitting favors a stellar mass normalization $\sim 3$ times higher than that given by the NSA catalogue. 
	
	Much of the difference between the observation required stellar mass normalization and that given by the NSA catalogue can be alleviated by using a more bottom-heavy initial mass function (IMF). The stellar mass of the NSA catalogue is derived by fitting stellar population templates to the broadband photometry data with a Chabrier initial mass function \citep[][]{Chabrier_2004}. Migrating the IMF from a Chabrier IMF to the Salpeter IMF~\citep{Salpeter_1955} on average can cause a 0.25 dex higher results of total stellar mass. Many recent observations show that the IMF may indeed vary as a function of velocity dispersion for the ETGs \citep[see a review of][]{Smith_2020}, and that the massive ETGs have a heavier IMF than that of Milky Way \citep{Auger_2010, Conroy_vanDokkum2012, Cappellari2012, Cappellari2013, Spiniello2014, LiHongyu2017,lushengdong2023b}. \citet{Schulz_Mandelbaum_Padmanabhan_2010} combined weak lensing and velocity dispersion observations to study the DM profiles of elliptical galaxies from the SDSS. They found a significant excess of mass in the interior compared to the prediction of the NFW model and this excess mass can be explained by the theoretical model of adiabatic contraction \citep[AC,][]{Gnedin_2004ApJ...616...16G} when stellar mass is obtained with a bottom-light Kroupa IMF~\citep{Kroupa2001MNRAS.322..231K}. In order to fully explain the observation without AC, the stellar masses would need to increase by a factor of two, meaning that a bottom-heavy Salpeter IMF would be required.
     In this work, we derive a $\alpha_{\rm nsa}=3.34_{-0.73}^{+0.52}$ and $\alpha_{\rm nsa}=2.87^{+1.12}_{-1.38}$ for the group and cluster bin respectively, which are slightly larger, but consistent with the value from pure stellar kinematic results of the MaNGA observation (vertical solid line in the histogram). 
 
    Given the condition of the data, in this paper, we did not consider the $M/L$ gradient of galaxies and assumed that the $M/L$ does not change with radius. However, if galaxies inherently have a stellar $M/L$ gradient (for example, \citet{Sonnenfeld_2018} using the Bayesian hierarchical modeling method identified a gradient in galaxy $M/L$), then our results on stellar mass and dark matter decomposition would be biased. If the stellar $M/L$ of a galaxy decreases with the galaxy's radius, then in the outer regions of the galaxy, our model would overestimate the dark matter mass fraction, consequently underestimating the dark matter density slope. Nevertheless, our article primarily focuses on central galaxies in galaxy groups, which are generally massive elliptical galaxies. According to \citet{lihongyu2018} and \citet{lushengdong2023}, these galaxies usually have a flat $M/L$ gradient.
	
	We finally explore whether the observational data can also be explained by using a steeper dark matter profile. If we fix the stellar mass normalization to the NSA value, the data requires a steep inner density profile for a gNFW dark matter model, with $\gamma_{\rm gnfw}=1.82_{-0.25}^{+0.15}$ , and a mass-weighted density slope $\overline{\gamma}_{\rm dm}(<R_{\rm e})=1.83_{-0.22}^{+0.13}$ for the group bin, higher than that predicted by the NFW model. The best-fit mass-weighted total density slope does not depend strongly on the choice of stellar component normalization, which is $\overline{\gamma}_{\rm tot}(<R_{\rm e})=2.12_{-0.09}^{+0.05}$ and $\overline{\gamma}_{\rm tot}(<R_{\rm e})=1.93_{-0.10}^{+0.06}$ for the two bins respectively. The total density slope is steeper than the values from TNG300 by about $2\sigma$ level.

	The next generation weak lensing survey, such as China Space Station Telescope (CSST)~\citep{zhanhu2011SSPMA..41.1441Z, Zhanhu2021} mission, and the Euclid~\citep{Euclid2011arXiv1110.3193L} mission will all provide a weak lensing source sample at least one magnitude larger than the current DECaLS sample. Combining the stellar kinematic data and these incoming weak lensing data may eventually break the degeneracy between stellar mass normalization and dark matter inner density slope, helping us understand the interplay between dark and light at the very center of dark matter haloes. 
	
	\section*{Acknowledgements}
      We acknowledge the support by National Key R\&D Program of China No. 2022YFF0503403, the support of National Nature Science Foundation of China (Nos 11988101,12022306), the support from the Ministry of Science and Technology of China (Nos. 2020SKA0110100), the science research grants from the China Manned Space Project (Nos. CMS-CSST-2021-B01, CMS-CSST-2021-A01), CAS Project for Young Scientists in Basic Research (No. YSBR-062), and the support from K.C.Wong Education Foundation. HYS acknowledges the support from NSFC of China under grant 11973070, Key Research Program of Frontier Sciences, CAS, Grant No. ZDBS-LY-7013 and Program of Shanghai Academic/Technology Research Leader. We acknowledge the support from the science research grants from the China Manned Space Project with NO. CMS-CSST-2021-A01, CMS-CSST-2021-A04. WWX acknowledges support from the National Science Foundation of China (11721303, 11890693, 12203063) and the National Key R$\&$D Program of China (2016YFA0400703). JY acknowledges the support from NSFC Grant No. 12203084, the China Postdoctoral Science Foundation Grant No. 2021T140451, and the Shanghai Post-doctoral Excellence Program Grant No. 2021419.
	
%%%%%%%%%%%%%%%%%%%%%%%%%%%%%%%%%%%%%%%%%%%%%%%%%%

%%%%%%%%%%%%%%%%%%%%%%%%%%%%%%%%%%%%%%%%%%%%%%%%%%
   \section{Data Availability}

   The catalogues of dynamical properties \citep{Zhu_2023_paperI} and stellar population properties \citep{lushengdong2023} are publicly available on the website of MaNGA DynPop (\url{https://manga-dynpop.github.io}). Other data underlying this article will be shared on a reasonable request to the authors.

	
	
	\bibliographystyle{mnras}
	\bibliography{moras} % if your bibtex file is called example.bib

\begin{thebibliography}{}
\makeatletter
\relax
\def\mn@urlcharsother{\let\do\@makeother \do\$\do\&\do\#\do\^\do\_\do\%\do\~}
\def\mn@doi{\begingroup\mn@urlcharsother \@ifnextchar [ {\mn@doi@}
  {\mn@doi@[]}}
\def\mn@doi@[#1]#2{\def\@tempa{#1}\ifx\@tempa\@empty \href
  {http://dx.doi.org/#2} {doi:#2}\else \href {http://dx.doi.org/#2} {#1}\fi
  \endgroup}
\def\mn@eprint#1#2{\mn@eprint@#1:#2::\@nil}
\def\mn@eprint@arXiv#1{\href {http://arxiv.org/abs/#1} {{\tt arXiv:#1}}}
\def\mn@eprint@dblp#1{\href {http://dblp.uni-trier.de/rec/bibtex/#1.xml}
  {dblp:#1}}
\def\mn@eprint@#1:#2:#3:#4\@nil{\def\@tempa {#1}\def\@tempb {#2}\def\@tempc
  {#3}\ifx \@tempc \@empty \let \@tempc \@tempb \let \@tempb \@tempa \fi \ifx
  \@tempb \@empty \def\@tempb {arXiv}\fi \@ifundefined
  {mn@eprint@\@tempb}{\@tempb:\@tempc}{\expandafter \expandafter \csname
  mn@eprint@\@tempb\endcsname \expandafter{\@tempc}}}

\bibitem[\protect\citeauthoryear{Abdurro'uf et~al.,}{Abdurro'uf
  et~al.}{2022}]{Abdurrouf2022}
Abdurro'uf et~al., 2022, \mn@doi [\apjs] {10.3847/1538-4365/ac4414}, 259, 35

\bibitem[\protect\citeauthoryear{Auger, Treu, Gavazzi, Bolton, Koopmans  \&
  Marshall}{Auger et~al.}{2010a}]{Auger_2010}
Auger M.~W.,  Treu T.,  Gavazzi R.,  Bolton A.~S.,  Koopmans L. V.~E.,
  Marshall P.~J.,  2010a, \mn@doi [\apj] {10.1088/2041-8205/721/2/l163}, 721,
  L163

\bibitem[\protect\citeauthoryear{{Auger}, {Treu}, {Bolton}, {Gavazzi},
  {Koopmans}, {Marshall}, {Moustakas}  \& {Burles}}{{Auger}
  et~al.}{2010b}]{Auger_2010b}
{Auger} M.~W.,  {Treu} T.,  {Bolton} A.~S.,  {Gavazzi} R.,  {Koopmans}
  L.~V.~E.,  {Marshall} P.~J.,  {Moustakas} L.~A.,   {Burles} S.,  2010b,
  \mn@doi [\apj] {10.1088/0004-637X/724/1/511}, \href
  {https://qa.adsabs.harvard.edu/abs/2010ApJ...724..511A} {724, 511}

\bibitem[\protect\citeauthoryear{Bellstedt et~al.,}{Bellstedt
  et~al.}{2018}]{Bellstedt2018}
Bellstedt S.,  et~al., 2018, \mn@doi [\mnras] {10.1093/mnras/sty456}, 476, 4543

\bibitem[\protect\citeauthoryear{Binney \& Tremaine}{Binney \&
  Tremaine}{2008}]{Binney2008}
Binney J.,  Tremaine S.,  2008, Galactic Dynamics: Second Edition.
Princeton University Press, Princeton, NJ, \url
  {https://books.google.co.uk/books?id=6mF4CKxlbLsC}

\bibitem[\protect\citeauthoryear{{Blanton} \& {Roweis}}{{Blanton} \&
  {Roweis}}{2007}]{Blanton2007}
{Blanton} M.~R.,  {Roweis} S.,  2007, \mn@doi [\aj] {10.1086/510127}, \href
  {https://ui.adsabs.harvard.edu/abs/2007AJ....133..734B} {133, 734}

\bibitem[\protect\citeauthoryear{Blanton, Kazin, Muna, Weaver  \&
  Price-Whelan}{Blanton et~al.}{2011}]{Blanton2011}
Blanton M.~R.,  Kazin E.,  Muna D.,  Weaver B.~A.,   Price-Whelan A.,  2011,
  \mn@doi [\aj] {10.1088/0004-6256/142/1/31}, 142, 31

\bibitem[\protect\citeauthoryear{Blumenthal, Faber, Flores  \&
  Primack}{Blumenthal et~al.}{1986}]{Blumenthal1986}
Blumenthal G.~R.,  Faber S.~M.,  Flores R.,   Primack J.~R.,  1986, \mn@doi
  [\apj] {10.1086/163867}, 301, 27

\bibitem[\protect\citeauthoryear{Bosma}{Bosma}{1978}]{Bosma1978}
Bosma A.,  1978, PhD thesis, Groningen Univ.

\bibitem[\protect\citeauthoryear{Brodie et~al.,}{Brodie
  et~al.}{2014}]{Brodie2014}
Brodie J.~P.,  et~al., 2014, \mn@doi [\apj] {10.1088/0004-637X/796/1/52}, \href
  {http://ads.ari.uni-heidelberg.de/abs/2014ApJ...796...52B} {796, 52}

\bibitem[\protect\citeauthoryear{{Bundy} et~al.,}{{Bundy}
  et~al.}{2015}]{Bundy2015}
{Bundy} K.,  et~al., 2015, \mn@doi [\apj] {10.1088/0004-637X/798/1/7}, \href
  {https://ui.adsabs.harvard.edu/abs/2015ApJ...798....7B} {798, 7}

\bibitem[\protect\citeauthoryear{{Cappellari}}{{Cappellari}}{2002}]{Cappellari2002}
{Cappellari} M.,  2002, \mn@doi [\mnras] {10.1046/j.1365-8711.2002.05412.x},
  \href {https://ui.adsabs.harvard.edu/abs/2002MNRAS.333..400C} {333, 400}

\bibitem[\protect\citeauthoryear{{Cappellari}}{{Cappellari}}{2008}]{Cappellari2008}
{Cappellari} M.,  2008, \mn@doi [\mnras] {10.1111/j.1365-2966.2008.13754.x},
  \href {https://ui.adsabs.harvard.edu/abs/2008MNRAS.390...71C} {390, 71}

\bibitem[\protect\citeauthoryear{Cappellari}{Cappellari}{2016}]{Cappellari2016}
Cappellari M.,  2016, \mn@doi [\araa] {10.1146/annurev-astro-082214-122432},
  \href {https://ui.adsabs.harvard.edu/abs/2016ARA%26A..54..597C} {54, 597}

\bibitem[\protect\citeauthoryear{{Cappellari}}{{Cappellari}}{2017}]{Cappellari2017}
{Cappellari} M.,  2017, \mn@doi [\mnras] {10.1093/mnras/stw3020}, \href
  {https://ui.adsabs.harvard.edu/abs/2017MNRAS.466..798C} {466, 798}

\bibitem[\protect\citeauthoryear{{Cappellari}}{{Cappellari}}{2020}]{Cappellari2020}
{Cappellari} M.,  2020, \mn@doi [\mnras] {10.1093/mnras/staa959}, \href
  {https://ui.adsabs.harvard.edu/abs/2020MNRAS.494.4819C} {494, 4819}

\bibitem[\protect\citeauthoryear{{Cappellari} \& {Copin}}{{Cappellari} \&
  {Copin}}{2003}]{Cappellari2003}
{Cappellari} M.,  {Copin} Y.,  2003, \mn@doi [\mnras]
  {10.1046/j.1365-8711.2003.06541.x}, \href
  {https://ui.adsabs.harvard.edu/abs/2003MNRAS.342..345C} {342, 345}

\bibitem[\protect\citeauthoryear{Cappellari et~al.,}{Cappellari
  et~al.}{2011}]{Cappellari2011}
Cappellari M.,  et~al., 2011, \mn@doi [\mnras]
  {10.1111/j.1365-2966.2010.18174.x}, \href
  {https://ui.adsabs.harvard.edu/abs/2011MNRAS.413..813C} {413, 813}

\bibitem[\protect\citeauthoryear{{Cappellari} et~al.,}{{Cappellari}
  et~al.}{2012}]{Cappellari2012}
{Cappellari} M.,  et~al., 2012, \mn@doi [\nat] {10.1038/nature10972}, \href
  {https://qa.adsabs.harvard.edu/abs/2012Natur.484..485C} {484, 485}

\bibitem[\protect\citeauthoryear{{Cappellari} et~al.,}{{Cappellari}
  et~al.}{2013}]{Cappellari2013}
{Cappellari} M.,  et~al., 2013, \mn@doi [\mnras] {10.1093/mnras/stt562}, \href
  {https://ui.adsabs.harvard.edu/abs/2013MNRAS.432.1709C} {432, 1709}

\bibitem[\protect\citeauthoryear{Cappellari et~al.,}{Cappellari
  et~al.}{2015}]{Cappellari2015}
Cappellari M.,  et~al., 2015, \mn@doi [\apjl] {10.1088/2041-8205/804/1/L21},
  \href {https://ui.adsabs.harvard.edu/abs/2015ApJ...804L..21C} {804, L21}

\bibitem[\protect\citeauthoryear{Chabrier}{Chabrier}{2003}]{Chabrier_2004}
Chabrier G.,  2003, \mn@doi [Publications of the Astronomical Society of the
  Pacific] {10.1086/376392}, 115, 763

\bibitem[\protect\citeauthoryear{{Conroy} \& {van Dokkum}}{{Conroy} \& {van
  Dokkum}}{2012}]{Conroy_vanDokkum2012}
{Conroy} C.,  {van Dokkum} P.~G.,  2012, \mn@doi [\apj]
  {10.1088/0004-637X/760/1/71}, \href
  {https://ui.adsabs.harvard.edu/abs/2012ApJ...760...71C} {760, 71}

\bibitem[\protect\citeauthoryear{{Coupon, J.} et~al.,}{{Coupon, J.}
  et~al.}{2012}]{Coupon_2012}
{Coupon, J.} et~al., 2012, \mn@doi [\aap] {10.1051/0004-6361/201117625}, 542,
  A5

\bibitem[\protect\citeauthoryear{Courteau et~al.,}{Courteau
  et~al.}{2014}]{Courteau2014}
Courteau S.,  et~al., 2014, \mn@doi [Reviews of Modern Physics]
  {10.1103/RevModPhys.86.47}, \href
  {https://ui.adsabs.harvard.edu/abs/2014RvMP...86...47C} {86, 47}

\bibitem[\protect\citeauthoryear{{Covone}, {Sereno}, {Kilbinger}  \&
  {Cardone}}{{Covone} et~al.}{2014}]{Covone2014}
{Covone} G.,  {Sereno} M.,  {Kilbinger} M.,   {Cardone} V.~F.,  2014, \mn@doi
  [\apjl] {10.1088/2041-8205/784/2/L25}, \href
  {https://qa.adsabs.harvard.edu/abs/2014ApJ...784L..25C} {784, L25}

\bibitem[\protect\citeauthoryear{{Dark Energy Survey Collaboration}
  et~al.,}{{Dark Energy Survey Collaboration} et~al.}{2016}]{DES2016}
{Dark Energy Survey Collaboration} et~al., 2016, \mn@doi [\mnras]
  {10.1093/mnras/stw641}, \href
  {https://ui.adsabs.harvard.edu/abs/2016MNRAS.460.1270D} {460, 1270}

\bibitem[\protect\citeauthoryear{Dey et~al.,}{Dey et~al.}{2019}]{Dey_2019}
Dey A.,  et~al., 2019, \mn@doi [\aj] {10.3847/1538-3881/ab089d}, 157, 168

\bibitem[\protect\citeauthoryear{{Dom{\'\i}nguez S{\'a}nchez}, {Margalef},
  {Bernardi}  \& {Huertas-Company}}{{Dom{\'\i}nguez S{\'a}nchez}
  et~al.}{2022}]{Dominguez-Sanchez2022}
{Dom{\'\i}nguez S{\'a}nchez} H.,  {Margalef} B.,  {Bernardi} M.,
  {Huertas-Company} M.,  2022, \mn@doi [\mnras] {10.1093/mnras/stab3089}, \href
  {https://ui.adsabs.harvard.edu/abs/2022MNRAS.509.4024D} {509, 4024}

\bibitem[\protect\citeauthoryear{Duffy, Schaye, Kay  \& Dalla~Vecchia}{Duffy
  et~al.}{2008}]{Duffy2008}
Duffy A.~R.,  Schaye J.,  Kay S.~T.,   Dalla~Vecchia C.,  2008, \mn@doi
  [\mnras: Letters] {10.1111/j.1745-3933.2008.00537.x}, 390, L64

\bibitem[\protect\citeauthoryear{{Duffy}, {Schaye}, {Kay}, {Dalla Vecchia},
  {Battye}  \& {Booth}}{{Duffy} et~al.}{2010}]{Duffy2010}
{Duffy} A.~R.,  {Schaye} J.,  {Kay} S.~T.,  {Dalla Vecchia} C.,  {Battye}
  R.~A.,   {Booth} C.~M.,  2010, \mn@doi [\mnras]
  {10.1111/j.1365-2966.2010.16613.x}, \href
  {https://ui.adsabs.harvard.edu/abs/2010MNRAS.405.2161D} {405, 2161}

\bibitem[\protect\citeauthoryear{Dutton \& Treu}{Dutton \&
  Treu}{2014}]{Dutton2014}
Dutton A.~A.,  Treu T.,  2014, \mn@doi [\mnras] {10.1093/mnras/stt2489}, \href
  {https://ui.adsabs.harvard.edu/abs/2014MNRAS.438.3594D} {438, 3594}

\bibitem[\protect\citeauthoryear{{Emsellem}, {Monnet}  \& {Bacon}}{{Emsellem}
  et~al.}{1994}]{Emsellem1994}
{Emsellem} E.,  {Monnet} G.,   {Bacon} R.,  1994, \aap, \href
  {https://ui.adsabs.harvard.edu/abs/1994A&A...285..723E} {285, 723}

\bibitem[\protect\citeauthoryear{Faber \& Gallagher}{Faber \&
  Gallagher}{1979}]{Faber1979}
Faber S.~M.,  Gallagher J.~S.,  1979, \mn@doi [\araa]
  {10.1146/annurev.aa.17.090179.001031}, \href
  {https://ui.adsabs.harvard.edu/abs/1979ARA%26A..17..135F} {17, 135}

\bibitem[\protect\citeauthoryear{{Foreman-Mackey}, {Hogg}, {Lang}  \&
  {Goodman}}{{Foreman-Mackey} et~al.}{2013}]{2013PASP..125..306F}
{Foreman-Mackey} D.,  {Hogg} D.~W.,  {Lang} D.,   {Goodman} J.,  2013, \mn@doi
  [\pasp] {10.1086/670067}, \href
  {https://ui.adsabs.harvard.edu/abs/2013PASP..125..306F} {125, 306}

\bibitem[\protect\citeauthoryear{{Forestell} \& {Gebhardt}}{{Forestell} \&
  {Gebhardt}}{2010}]{Forestell_2010ApJ...716..370F}
{Forestell} A.~D.,  {Gebhardt} K.,  2010, \mn@doi [\apj]
  {10.1088/0004-637X/716/1/370}, \href
  {https://ui.adsabs.harvard.edu/abs/2010ApJ...716..370F} {716, 370}

\bibitem[\protect\citeauthoryear{Frenk \& White}{Frenk \&
  White}{2012}]{Frenk_and_White_2012}
Frenk C.,  White S.,  2012, \mn@doi [Annalen der Physik]
  {https://doi.org/10.1002/andp.201200212}, 524, 507

\bibitem[\protect\citeauthoryear{{Gao} \& {White}}{{Gao} \&
  {White}}{2006}]{Gao2006MNRAS.373...65G}
{Gao} L.,  {White} S. D.~M.,  2006, \mn@doi [\mnras]
  {10.1111/j.1365-2966.2006.11048.x}, \href
  {https://ui.adsabs.harvard.edu/abs/2006MNRAS.373...65G} {373, 65}

\bibitem[\protect\citeauthoryear{{Gao}, {Frenk}, {Boylan-Kolchin}, {Jenkins},
  {Springel}  \& {White}}{{Gao} et~al.}{2011}]{Gao2011}
{Gao} L.,  {Frenk} C.~S.,  {Boylan-Kolchin} M.,  {Jenkins} A.,  {Springel} V.,
   {White} S.~D.~M.,  2011, \mn@doi [\mnras]
  {10.1111/j.1365-2966.2010.17601.x}, \href
  {https://ui.adsabs.harvard.edu/abs/2011MNRAS.410.2309G} {410, 2309}

\bibitem[\protect\citeauthoryear{{Genel} et~al.,}{{Genel}
  et~al.}{2018}]{Genel2018MNRAS.474.3976G}
{Genel} S.,  et~al., 2018, \mn@doi [\mnras] {10.1093/mnras/stx3078}, \href
  {https://ui.adsabs.harvard.edu/abs/2018MNRAS.474.3976G} {474, 3976}

\bibitem[\protect\citeauthoryear{Giblin et~al.,}{Giblin
  et~al.}{2021}]{Giblin2021}
Giblin B.,  et~al., 2021, \mn@doi [\aap] {10.1051/0004-6361/202038850}, 645,
  A105

\bibitem[\protect\citeauthoryear{{Giocoli}, {Meneghetti}, {Ettori}  \&
  {Moscardini}}{{Giocoli} et~al.}{2012}]{Giocoli2012}
{Giocoli} C.,  {Meneghetti} M.,  {Ettori} S.,   {Moscardini} L.,  2012, \mn@doi
  [\mnras] {10.1111/j.1365-2966.2012.21743.x}, \href
  {https://ui.adsabs.harvard.edu/abs/2012MNRAS.426.1558G} {426, 1558}

\bibitem[\protect\citeauthoryear{Gnedin, Kravtsov, Klypin  \& Nagai}{Gnedin
  et~al.}{2004a}]{Gnedin2004}
Gnedin O.~Y.,  Kravtsov A.~V.,  Klypin A.~A.,   Nagai D.,  2004a, \mn@doi
  [\apj] {10.1086/424914}, 616, 12

\bibitem[\protect\citeauthoryear{{Gnedin}, {Kravtsov}, {Klypin}  \&
  {Nagai}}{{Gnedin} et~al.}{2004b}]{Gnedin_2004ApJ...616...16G}
{Gnedin} O.~Y.,  {Kravtsov} A.~V.,  {Klypin} A.~A.,   {Nagai} D.,  2004b,
  \mn@doi [\apj] {10.1086/424914}, \href
  {https://ui.adsabs.harvard.edu/abs/2004ApJ...616...16G} {616, 16}

\bibitem[\protect\citeauthoryear{{Gustafsson}, {Fairbairn}  \&
  {Sommer-Larsen}}{{Gustafsson} et~al.}{2006}]{Gustafsson2006}
{Gustafsson} M.,  {Fairbairn} M.,   {Sommer-Larsen} J.,  2006, \mn@doi [\prd]
  {10.1103/PhysRevD.74.123522}, \href
  {https://ui.adsabs.harvard.edu/abs/2006PhRvD..74l3522G} {74, 123522}

\bibitem[\protect\citeauthoryear{He et~al.,}{He et~al.}{2020}]{He2020}
He Q.,  et~al., 2020, \mn@doi [\mnras] {10.1093/mnras/staa1769}, 496, 4717

\bibitem[\protect\citeauthoryear{{Heymans} et~al.,}{{Heymans}
  et~al.}{2012}]{Heymans2012MNRAS.427..146H}
{Heymans} C.,  et~al., 2012, \mn@doi [\mnras]
  {10.1111/j.1365-2966.2012.21952.x}, \href
  {https://ui.adsabs.harvard.edu/abs/2012MNRAS.427..146H} {427, 146}

\bibitem[\protect\citeauthoryear{{Hildebrandt} et~al.}{{Hildebrandt}
  et~al.}{2017}]{Hildebrandt2017}
{Hildebrandt} H.,  et~al., 2017, \mn@doi [\mnras] {10.1093/mnras/stw2805},
  \href {http://adsabs.harvard.edu/abs/2017MNRAS.465.1454H} {465, 1454}

\bibitem[\protect\citeauthoryear{{Hoekstra}, {Franx}, {Kuijken}  \& {van
  Dokkum}}{{Hoekstra} et~al.}{2002}]{Hoekstra2002}
{Hoekstra} H.,  {Franx} M.,  {Kuijken} K.,   {van Dokkum} P.~G.,  2002, \mn@doi
  [\mnras] {10.1046/j.1365-8711.2002.05479.x}, \href
  {https://ui.adsabs.harvard.edu/abs/2002MNRAS.333..911H} {333, 911}

\bibitem[\protect\citeauthoryear{{Klypin}, {Yepes}, {Gottl{\"o}ber}, {Prada}
  \& {He{\ss}}}{{Klypin} et~al.}{2016}]{Klypin2016}
{Klypin} A.,  {Yepes} G.,  {Gottl{\"o}ber} S.,  {Prada} F.,   {He{\ss}} S.,
  2016, \mn@doi [\mnras] {10.1093/mnras/stw248}, \href
  {https://qa.adsabs.harvard.edu/abs/2016MNRAS.457.4340K} {457, 4340}

\bibitem[\protect\citeauthoryear{{Koopmans} et~al.,}{{Koopmans}
  et~al.}{2009}]{Koopmans2009}
{Koopmans} L.~V.~E.,  et~al., 2009, \mn@doi [\apjl]
  {10.1088/0004-637X/703/1/L51}, \href
  {https://ui.adsabs.harvard.edu/abs/2009ApJ...703L..51K} {703, L51}

\bibitem[\protect\citeauthoryear{{Kroupa}}{{Kroupa}}{2001}]{Kroupa2001MNRAS.322..231K}
{Kroupa} P.,  2001, \mn@doi [\mnras] {10.1046/j.1365-8711.2001.04022.x}, \href
  {https://ui.adsabs.harvard.edu/abs/2001MNRAS.322..231K} {322, 231}

\bibitem[\protect\citeauthoryear{{Lablanche} et~al.,}{{Lablanche}
  et~al.}{2012}]{Lablanche2012}
{Lablanche} P.-Y.,  et~al., 2012, \mn@doi [\mnras]
  {10.1111/j.1365-2966.2012.21343.x}, \href
  {https://ui.adsabs.harvard.edu/abs/2012MNRAS.424.1495L} {424, 1495}

\bibitem[\protect\citeauthoryear{{Lang}, {Hogg}  \& {Schlegel}}{{Lang}
  et~al.}{2016}]{Lang2016AJ....151...36L}
{Lang} D.,  {Hogg} D.~W.,   {Schlegel} D.~J.,  2016, \mn@doi [\aj]
  {10.3847/0004-6256/151/2/36}, \href
  {https://ui.adsabs.harvard.edu/abs/2016AJ....151...36L} {151, 36}

\bibitem[\protect\citeauthoryear{{Laureijs} et~al.,}{{Laureijs}
  et~al.}{2011}]{Euclid2011arXiv1110.3193L}
{Laureijs} R.,  et~al., 2011, \mn@doi [arXiv e-prints]
  {10.48550/arXiv.1110.3193}, \href
  {https://ui.adsabs.harvard.edu/abs/2011arXiv1110.3193L} {p. arXiv:1110.3193}

\bibitem[\protect\citeauthoryear{{Law} et~al.,}{{Law} et~al.}{2016}]{Law2016}
{Law} D.~R.,  et~al., 2016, \mn@doi [\aj] {10.3847/0004-6256/152/4/83}, \href
  {https://ui.adsabs.harvard.edu/abs/2016AJ....152...83L} {152, 83}

\bibitem[\protect\citeauthoryear{{Leauthaud} et~al.,}{{Leauthaud}
  et~al.}{2010}]{Leauthaud2010ApJ...709...97L}
{Leauthaud} A.,  et~al., 2010, \mn@doi [\apj] {10.1088/0004-637X/709/1/97},
  \href {https://ui.adsabs.harvard.edu/abs/2010ApJ...709...97L} {709, 97}

\bibitem[\protect\citeauthoryear{{Li}, {Li}, {Mao}, {Xu}, {Long}  \&
  {Emsellem}}{{Li} et~al.}{2016}]{Lihongyu2016}
{Li} H.,  {Li} R.,  {Mao} S.,  {Xu} D.,  {Long} R.~J.,   {Emsellem} E.,  2016,
  \mn@doi [\mnras] {10.1093/mnras/stv2565}, \href
  {https://ui.adsabs.harvard.edu/abs/2016MNRAS.455.3680L} {455, 3680}

\bibitem[\protect\citeauthoryear{{Li} et~al.,}{{Li}
  et~al.}{2017}]{LiHongyu2017}
{Li} H.,  et~al., 2017, \mn@doi [\apj] {10.3847/1538-4357/aa662a}, \href
  {https://ui.adsabs.harvard.edu/abs/2017ApJ...838...77L} {838, 77}

\bibitem[\protect\citeauthoryear{{Li} et~al.,}{{Li}
  et~al.}{2018a}]{lihongyu2018}
{Li} H.,  et~al., 2018a, \mn@doi [\mnras] {10.1093/mnras/sty334}, \href
  {https://ui.adsabs.harvard.edu/abs/2018MNRAS.476.1765L} {476, 1765}

\bibitem[\protect\citeauthoryear{Li, Shu  \& Wang}{Li
  et~al.}{2018b}]{Lirui2018}
Li R.,  Shu Y.,   Wang J.,  2018b, \mn@doi [\mnras] {10.1093/mnras/sty1813},
  480, 431

\bibitem[\protect\citeauthoryear{Li et~al.,}{Li et~al.}{2019}]{liran_2019}
Li R.,  et~al., 2019, \mn@doi [\mnras] {10.1093/mnras/stz2565}, 490, 2124

\bibitem[\protect\citeauthoryear{{Lu}, {Zhu}, {Cappellari}, {Li}, {Mao}  \&
  {Xu}}{{Lu} et~al.}{2023a}]{lushengdong2023}
{Lu} S.,  {Zhu} K.,  {Cappellari} M.,  {Li} R.,  {Mao} S.,   {Xu} D.,  2023a,
  \mn@doi [\mnras] {10.1093/mnras/stad2732}, \href
  {https://ui.adsabs.harvard.edu/abs/2023MNRAS.tmp.2611L} {}

\bibitem[\protect\citeauthoryear{{Lu}, {Zhu}, {Cappellari}, {Li}, {Mao}  \&
  {Xu}}{{Lu} et~al.}{2023b}]{lushengdong2023b}
{Lu} S.,  {Zhu} K.,  {Cappellari} M.,  {Li} R.,  {Mao} S.,   {Xu} D.,  2023b,
  \mn@doi [arXiv e-prints] {10.48550/arXiv.2309.12395}, \href
  {https://ui.adsabs.harvard.edu/abs/2023arXiv230912395L} {p. arXiv:2309.12395}

\bibitem[\protect\citeauthoryear{Mandelbaum et~al.,}{Mandelbaum
  et~al.}{2005}]{Mandelbaum2005boostfactor}
Mandelbaum R.,  et~al., 2005, \mn@doi [\mnras]
  {10.1111/j.1365-2966.2005.09282.x}, 361, 1287

\bibitem[\protect\citeauthoryear{{Mandelbaum}, {Seljak}, {Cool}, {Blanton},
  {Hirata}  \& {Brinkmann}}{{Mandelbaum} et~al.}{2006}]{Mandelbaum2006_372_758}
{Mandelbaum} R.,  {Seljak} U.,  {Cool} R.~J.,  {Blanton} M.,  {Hirata} C.~M.,
  {Brinkmann} J.,  2006, \mn@doi [\mnras] {10.1111/j.1365-2966.2006.10906.x},
  \href {https://ui.adsabs.harvard.edu/abs/2006MNRAS.372..758M} {372, 758}

\bibitem[\protect\citeauthoryear{{Mandelbaum}, {Seljak}  \&
  {Hirata}}{{Mandelbaum} et~al.}{2008}]{Mandelbaum2008}
{Mandelbaum} R.,  {Seljak} U.,   {Hirata} C.~M.,  2008, \mn@doi [\jcap]
  {10.1088/1475-7516/2008/08/006}, \href
  {https://qa.adsabs.harvard.edu/abs/2008JCAP...08..006M} {2008, 006}

\bibitem[\protect\citeauthoryear{Martinsson, Verheijen, Westfall, Bershady,
  Andersen  \& Swaters}{Martinsson et~al.}{2013}]{Martinsson2013}
Martinsson T. P.~K.,  Verheijen M. A.~W.,  Westfall K.~B.,  Bershady M.~A.,
  Andersen D.~R.,   Swaters R.~A.,  2013, \mn@doi [\aap]
  {10.1051/0004-6361/201321390}, 557, A131

\bibitem[\protect\citeauthoryear{{Meisner}, {Lang}  \& {Schlegel}}{{Meisner}
  et~al.}{2017}]{Meisner2017}
{Meisner} A.~M.,  {Lang} D.,   {Schlegel} D.~J.,  2017, \mn@doi [\aj]
  {10.3847/1538-3881/aa894e}, \href
  {https://ui.adsabs.harvard.edu/abs/2017AJ....154..161M} {154, 161}

\bibitem[\protect\citeauthoryear{{Merten} et~al.,}{{Merten}
  et~al.}{2015}]{Merten2015}
{Merten} J.,  et~al., 2015, \mn@doi [\apj] {10.1088/0004-637X/806/1/4}, \href
  {https://qa.adsabs.harvard.edu/abs/2015ApJ...806....4M} {806, 4}

\bibitem[\protect\citeauthoryear{Miller, Kitching, Heymans, Heavens  \& {Van
  Waerbeke}}{Miller et~al.}{2007}]{Miller2007}
Miller L.,  Kitching T.~D.,  Heymans C.,  Heavens A.~F.,   {Van Waerbeke} L.,
  2007, \mn@doi [\mnras] {10.1111/j.1365-2966.2007.12363.x}, 382, 315

\bibitem[\protect\citeauthoryear{Miller et~al.,}{Miller
  et~al.}{2013}]{Miller2013}
Miller L.,  et~al., 2013, \mn@doi [\mnras] {10.1093/mnras/sts454}, 429, 2858

\bibitem[\protect\citeauthoryear{{Miralda-Escude}}{{Miralda-Escude}}{1991}]{Miralda-Escude1991ApJ...370....1M}
{Miralda-Escude} J.,  1991, \mn@doi [\apj] {10.1086/169789}, \href
  {https://ui.adsabs.harvard.edu/abs/1991ApJ...370....1M} {370, 1}

\bibitem[\protect\citeauthoryear{{Moraes} et~al.,}{{Moraes}
  et~al.}{2014}]{Moraes2014}
{Moraes} B.,  et~al., 2014, in Revista Mexicana de Astronomia y Astrofisica
  Conference Series. pp 202--203

\bibitem[\protect\citeauthoryear{{Morganti}, {Gerhard}, {Coccato},
  {Martinez-Valpuesta}  \& {Arnaboldi}}{{Morganti}
  et~al.}{2013}]{Morganti_2013MNRAS.431.3570M}
{Morganti} L.,  {Gerhard} O.,  {Coccato} L.,  {Martinez-Valpuesta} I.,
  {Arnaboldi} M.,  2013, \mn@doi [\mnras] {10.1093/mnras/stt442}, \href
  {https://ui.adsabs.harvard.edu/abs/2013MNRAS.431.3570M} {431, 3570}

\bibitem[\protect\citeauthoryear{{Napolitano}, {Pota}, {Romanowsky}, {Forbes},
  {Brodie}  \& {Foster}}{{Napolitano}
  et~al.}{2014}]{Napolitano_2014MNRAS.439..659N}
{Napolitano} N.~R.,  {Pota} V.,  {Romanowsky} A.~J.,  {Forbes} D.~A.,  {Brodie}
  J.~P.,   {Foster} C.,  2014, \mn@doi [\mnras] {10.1093/mnras/stt2484}, \href
  {https://ui.adsabs.harvard.edu/abs/2014MNRAS.439..659N} {439, 659}

\bibitem[\protect\citeauthoryear{{Navarro}, {Eke}  \& {Frenk}}{{Navarro}
  et~al.}{1996a}]{NavarroEkeFrenk1996}
{Navarro} J.~F.,  {Eke} V.~R.,   {Frenk} C.~S.,  1996a, \mn@doi [\mnras]
  {10.1093/mnras/283.3.L72}, \href
  {https://ui.adsabs.harvard.edu/abs/1996MNRAS.283L..72N} {283, L72}

\bibitem[\protect\citeauthoryear{{Navarro}, {Frenk}  \& {White}}{{Navarro}
  et~al.}{1996b}]{Navarro1996}
{Navarro} J.~F.,  {Frenk} C.~S.,   {White} S. D.~M.,  1996b, \mn@doi [\apj]
  {10.1086/177173}, \href
  {https://ui.adsabs.harvard.edu/abs/1996ApJ...462..563N} {462, 563}

\bibitem[\protect\citeauthoryear{Navarro, Frenk  \& White}{Navarro
  et~al.}{1997}]{Navarro_1997}
Navarro J.~F.,  Frenk C.~S.,   White S. D.~M.,  1997, \mn@doi [\apj]
  {10.1086/304888}, 490, 493

\bibitem[\protect\citeauthoryear{Newman, Treu, Ellis, Sand, Nipoti, Richard  \&
  Jullo}{Newman et~al.}{2013}]{Newman_2013}
Newman A.~B.,  Treu T.,  Ellis R.~S.,  Sand D.~J.,  Nipoti C.,  Richard J.,
  Jullo E.,  2013, \mn@doi [The Astrophysical Journal]
  {10.1088/0004-637X/765/1/24}, 765, 24

\bibitem[\protect\citeauthoryear{Newman, Ellis  \& Treu}{Newman
  et~al.}{2015}]{Newman_2015}
Newman A.~B.,  Ellis R.~S.,   Treu T.,  2015, \mn@doi [\apj]
  {10.1088/0004-637x/814/1/26}, 814, 26

\bibitem[\protect\citeauthoryear{Nitschai, Eilers, Neumayer, Cappellari  \&
  Rix}{Nitschai et~al.}{2021}]{Nitschai2021}
Nitschai M.~S.,  Eilers A.-C.,  Neumayer N.,  Cappellari M.,   Rix H.-W.,
  2021, \mn@doi [The Astrophysical Journal] {10.3847/1538-4357/ac04b5}, 916,
  112

\bibitem[\protect\citeauthoryear{{Oaxaca Wright} \& {Brainerd}}{{Oaxaca Wright}
  \& {Brainerd}}{1999}]{Wright1999astro.ph..8213O}
{Oaxaca Wright} C.,  {Brainerd} T.~G.,  1999, \mn@doi [arXiv e-prints]
  {10.48550/arXiv.astro-ph/9908213}, \href
  {https://ui.adsabs.harvard.edu/abs/1999astro.ph..8213O} {pp
  astro--ph/9908213}

\bibitem[\protect\citeauthoryear{{Oguri} \& {Hamana}}{{Oguri} \&
  {Hamana}}{2011}]{Oguri_and_Hamana_2011}
{Oguri} M.,  {Hamana} T.,  2011, \mn@doi [\mnras]
  {10.1111/j.1365-2966.2011.18481.x}, \href
  {https://ui.adsabs.harvard.edu/abs/2011MNRAS.414.1851O} {414, 1851}

\bibitem[\protect\citeauthoryear{{Phriksee}, {Jullo}, {Limousin}, {Shan},
  {Finoguenov}, {Komonjinda}, {Wannawichian}  \& {Sawangwit}}{{Phriksee}
  et~al.}{2020a}]{Phriksee2020MNRAS.491.1643P}
{Phriksee} A.,  {Jullo} E.,  {Limousin} M.,  {Shan} H.,  {Finoguenov} A.,
  {Komonjinda} S.,  {Wannawichian} S.,   {Sawangwit} U.,  2020a, \mn@doi
  [\mnras] {10.1093/mnras/stz3049}, \href
  {https://ui.adsabs.harvard.edu/abs/2020MNRAS.491.1643P} {491, 1643}

\bibitem[\protect\citeauthoryear{{Phriksee}, {Jullo}, {Limousin}, {Shan},
  {Finoguenov}, {Komonjinda}, {Wannawichian}  \& {Sawangwit}}{{Phriksee}
  et~al.}{2020b}]{Phriksee2020}
{Phriksee} A.,  {Jullo} E.,  {Limousin} M.,  {Shan} H.,  {Finoguenov} A.,
  {Komonjinda} S.,  {Wannawichian} S.,   {Sawangwit} U.,  2020b, \mn@doi
  [\mnras] {10.1093/mnras/stz3049}, \href
  {https://ui.adsabs.harvard.edu/abs/2020MNRAS.491.1643P} {491, 1643}

\bibitem[\protect\citeauthoryear{{Planck Collaboration} et~al.,}{{Planck
  Collaboration} et~al.}{2016}]{Planck_Collaboration2016}
{Planck Collaboration} et~al., 2016, \mn@doi [\aap]
  {10.1051/0004-6361/201525830}, \href
  {https://ui.adsabs.harvard.edu/abs/2016A&A...594A..13P} {594, A13}

\bibitem[\protect\citeauthoryear{{Poci}, {Cappellari}  \& {McDermid}}{{Poci}
  et~al.}{2017}]{Poci2017}
{Poci} A.,  {Cappellari} M.,   {McDermid} R.~M.,  2017, \mn@doi [\mnras]
  {10.1093/mnras/stx101}, \href
  {https://ui.adsabs.harvard.edu/abs/2017MNRAS.467.1397P} {467, 1397}

\bibitem[\protect\citeauthoryear{{Pontzen} \& {Governato}}{{Pontzen} \&
  {Governato}}{2012}]{Pontzen2012}
{Pontzen} A.,  {Governato} F.,  2012, \mn@doi [\mnras]
  {10.1111/j.1365-2966.2012.20571.x}, \href
  {https://ui.adsabs.harvard.edu/abs/2012MNRAS.421.3464P} {421, 3464}

\bibitem[\protect\citeauthoryear{{Read} \& {Gilmore}}{{Read} \&
  {Gilmore}}{2005}]{Read2005}
{Read} J.~I.,  {Gilmore} G.,  2005, \mn@doi [\mnras]
  {10.1111/j.1365-2966.2004.08424.x}, \href
  {https://ui.adsabs.harvard.edu/abs/2005MNRAS.356..107R} {356, 107}

\bibitem[\protect\citeauthoryear{Rubin, Ford  \& Thonnard}{Rubin
  et~al.}{1980}]{Rubin1980}
Rubin V.~C.,  Ford W. K.~J.,   Thonnard N.,  1980, \mn@doi [\apj]
  {10.1086/158003}, \href
  {https://ui.adsabs.harvard.edu/abs/1980ApJ...238..471R} {238, 471}

\bibitem[\protect\citeauthoryear{{Salpeter}}{{Salpeter}}{1955}]{Salpeter_1955}
{Salpeter} E.~E.,  1955, \mn@doi [\apj] {10.1086/145971}, \href
  {https://ui.adsabs.harvard.edu/abs/1955ApJ...121..161S} {121, 161}

\bibitem[\protect\citeauthoryear{{S{\'a}nchez-Bl{\'a}zquez}
  et~al.,}{{S{\'a}nchez-Bl{\'a}zquez} et~al.}{2006}]{Sanchez-Blazquez2006}
{S{\'a}nchez-Bl{\'a}zquez} P.,  et~al., 2006, \mn@doi [\mnras]
  {10.1111/j.1365-2966.2006.10699.x}, \href
  {https://ui.adsabs.harvard.edu/abs/2006MNRAS.371..703S} {371, 703}

\bibitem[\protect\citeauthoryear{{Sartoris} et~al.,}{{Sartoris}
  et~al.}{2020}]{Sartoris2020}
{Sartoris} B.,  et~al., 2020, \mn@doi [\aap] {10.1051/0004-6361/202037521},
  \href {https://qa.adsabs.harvard.edu/abs/2020A&A...637A..34S} {637, A34}

\bibitem[\protect\citeauthoryear{{Schaller} et~al.,}{{Schaller}
  et~al.}{2015}]{Schaller2015}
{Schaller} M.,  et~al., 2015, \mn@doi [\mnras] {10.1093/mnras/stv1341}, \href
  {https://ui.adsabs.harvard.edu/abs/2015MNRAS.452..343S} {452, 343}

\bibitem[\protect\citeauthoryear{Schulz, Mandelbaum  \& Padmanabhan}{Schulz
  et~al.}{2010}]{Schulz_Mandelbaum_Padmanabhan_2010}
Schulz A.~E.,  Mandelbaum R.,   Padmanabhan N.,  2010, \mn@doi [\mnras]
  {10.1111/j.1365-2966.2010.17207.x}, 408, 1463

\bibitem[\protect\citeauthoryear{Serra, Oosterloo, Cappellari, den Heijer  \&
  J{\'o}zsa}{Serra et~al.}{2016}]{Serra2016}
Serra P.,  Oosterloo T.,  Cappellari M.,  den Heijer M.,   J{\'o}zsa G. I.~G.,
  2016, \mn@doi [\mnras] {10.1093/mnras/stw1010}, \href
  {https://ui.adsabs.harvard.edu/abs/2016MNRAS.460.1382S} {460, 1382}

\bibitem[\protect\citeauthoryear{Shan et~al.,}{Shan et~al.}{2017}]{Shan2017}
Shan H.,  et~al., 2017, \apj, 840, 104

\bibitem[\protect\citeauthoryear{{Skrutskie} et~al.,}{{Skrutskie}
  et~al.}{2006}]{Skrutskie2006}
{Skrutskie} M.~F.,  et~al., 2006, \mn@doi [\aj] {10.1086/498708}, \href
  {https://ui.adsabs.harvard.edu/abs/2006AJ....131.1163S} {131, 1163}

\bibitem[\protect\citeauthoryear{{Smee} et~al.,}{{Smee}
  et~al.}{2013}]{Smee_2013}
{Smee} S.~A.,  et~al., 2013, \mn@doi [\aj] {10.1088/0004-6256/146/2/32}, \href
  {https://ui.adsabs.harvard.edu/abs/2013AJ....146...32S} {146, 32}

\bibitem[\protect\citeauthoryear{Smith}{Smith}{2020}]{Smith_2020}
Smith R.~J.,  2020, \mn@doi [\araa] {10.1146/annurev-astro-032620-020217}, 58,
  577

\bibitem[\protect\citeauthoryear{Sonnenfeld \& Leauthaud}{Sonnenfeld \&
  Leauthaud}{2018}]{Sonnenfeld_and_Leauthaud_2018}
Sonnenfeld A.,  Leauthaud A.,  2018, \mn@doi [\mnras] {10.1093/mnras/sty935},
  477, 5460

\bibitem[\protect\citeauthoryear{{Sonnenfeld}, {Treu}, {Gavazzi}, {Marshall},
  {Auger}, {Suyu}, {Koopmans}  \& {Bolton}}{{Sonnenfeld}
  et~al.}{2012}]{Sonnenfeld2012}
{Sonnenfeld} A.,  {Treu} T.,  {Gavazzi} R.,  {Marshall} P.~J.,  {Auger} M.~W.,
  {Suyu} S.~H.,  {Koopmans} L.~V.~E.,   {Bolton} A.~S.,  2012, \mn@doi [\apj]
  {10.1088/0004-637X/752/2/163}, \href
  {https://qa.adsabs.harvard.edu/abs/2012ApJ...752..163S} {752, 163}

\bibitem[\protect\citeauthoryear{Sonnenfeld, Leauthaud, Auger, Gavazzi, Treu,
  More  \& Komiyama}{Sonnenfeld et~al.}{2018}]{Sonnenfeld_2018}
Sonnenfeld A.,  Leauthaud A.,  Auger M.~W.,  Gavazzi R.,  Treu T.,  More S.,
  Komiyama Y.,  2018, \mn@doi [\mnras] {10.1093/mnras/sty2262}, 481, 164

\bibitem[\protect\citeauthoryear{{Spiniello}, {Trager}, {Koopmans}  \&
  {Conroy}}{{Spiniello} et~al.}{2014}]{Spiniello2014}
{Spiniello} C.,  {Trager} S.,  {Koopmans} L. V.~E.,   {Conroy} C.,  2014,
  \mn@doi [\mnras] {10.1093/mnras/stt2282}, \href
  {https://ui.adsabs.harvard.edu/abs/2014MNRAS.438.1483S} {438, 1483}

\bibitem[\protect\citeauthoryear{{Springel} et~al.,}{{Springel}
  et~al.}{2008}]{Springel2008}
{Springel} V.,  et~al., 2008, \mn@doi [\mnras]
  {10.1111/j.1365-2966.2008.14066.x}, \href
  {https://ui.adsabs.harvard.edu/abs/2008MNRAS.391.1685S} {391, 1685}

\bibitem[\protect\citeauthoryear{Tinker, Robertson, Kravtsov, Klypin, Warren,
  Yepes  \& Gottl{\"{o}}ber}{Tinker et~al.}{2010}]{Tinker2010}
Tinker J.~L.,  Robertson B.~E.,  Kravtsov A.~V.,  Klypin A.,  Warren M.~S.,
  Yepes G.,   Gottl{\"{o}}ber S.,  2010, \mn@doi [\apj]
  {10.1088/0004-637X/724/2/878}, 724, 878

\bibitem[\protect\citeauthoryear{Treu}{Treu}{2010}]{Treu2010}
Treu T.,  2010, \mn@doi [\araa] {10.1146/annurev-astro-081309-130924}, \href
  {https://ui.adsabs.harvard.edu/abs/2010ARA%26A..48...87T} {48, 87}

\bibitem[\protect\citeauthoryear{{Umetsu} et~al.,}{{Umetsu}
  et~al.}{2014}]{Umetsu2014}
{Umetsu} K.,  et~al., 2014, \mn@doi [\apj] {10.1088/0004-637X/795/2/163}, \href
  {https://qa.adsabs.harvard.edu/abs/2014ApJ...795..163U} {795, 163}

\bibitem[\protect\citeauthoryear{Wang et~al.,}{Wang et~al.}{2018}]{Wangcx_2018}
Wang C.,  et~al., 2018, \mn@doi [\mnras] {10.1093/mnras/sty073}, 475, 4020

\bibitem[\protect\citeauthoryear{{Weijmans}, {Krajnovi{\'c}}, {van de Ven},
  {Oosterloo}, {Morganti}  \& {de Zeeuw}}{{Weijmans}
  et~al.}{2008}]{Weijmans_2008MNRAS.383.1343W}
{Weijmans} A.-M.,  {Krajnovi{\'c}} D.,  {van de Ven} G.,  {Oosterloo} T.~A.,
  {Morganti} R.,   {de Zeeuw} P.~T.,  2008, \mn@doi [\mnras]
  {10.1111/j.1365-2966.2007.12680.x}, \href
  {https://ui.adsabs.harvard.edu/abs/2008MNRAS.383.1343W} {383, 1343}

\bibitem[\protect\citeauthoryear{{Weijmans} et~al.,}{{Weijmans}
  et~al.}{2009}]{Weijmans_2009MNRAS.398..561W}
{Weijmans} A.-M.,  et~al., 2009, \mn@doi [\mnras]
  {10.1111/j.1365-2966.2009.15134.x}, \href
  {https://ui.adsabs.harvard.edu/abs/2009MNRAS.398..561W} {398, 561}

\bibitem[\protect\citeauthoryear{{Westfall} et~al.,}{{Westfall}
  et~al.}{2019}]{Westfall2019}
{Westfall} K.~B.,  et~al., 2019, \mn@doi [\aj] {10.3847/1538-3881/ab44a2},
  \href {https://ui.adsabs.harvard.edu/abs/2019AJ....158..231W} {158, 231}

\bibitem[\protect\citeauthoryear{{Wilson}, {Kaiser}, {Luppino}  \&
  {Cowie}}{{Wilson} et~al.}{2001}]{Wilson2001ApJ...555..572W}
{Wilson} G.,  {Kaiser} N.,  {Luppino} G.~A.,   {Cowie} L.~L.,  2001, \mn@doi
  [\apj] {10.1086/321441}, \href
  {https://ui.adsabs.harvard.edu/abs/2001ApJ...555..572W} {555, 572}

\bibitem[\protect\citeauthoryear{Xu et~al.,}{Xu et~al.}{2021}]{Xu_2021}
Xu W.,  et~al., 2021, \mn@doi [The Astrophysical Journal]
  {10.3847/1538-4357/ac1b9e}, 922, 162

\bibitem[\protect\citeauthoryear{{Yan} et~al.,}{{Yan} et~al.}{2016}]{Yan2016}
{Yan} R.,  et~al., 2016, \mn@doi [\aj] {10.3847/0004-6256/151/1/8}, \href
  {https://ui.adsabs.harvard.edu/abs/2016AJ....151....8Y} {151, 8}

\bibitem[\protect\citeauthoryear{{Yang}, {Mo}, {van den Bosch}, {Weinmann},
  {Li}  \& {Jing}}{{Yang} et~al.}{2005}]{Yang2005}
{Yang} X.,  {Mo} H.~J.,  {van den Bosch} F.~C.,  {Weinmann} S.~M.,  {Li} C.,
  {Jing} Y.~P.,  2005, \mn@doi [\mnras] {10.1111/j.1365-2966.2005.09351.x},
  \href {http://adsabs.harvard.edu/abs/2005MNRAS.362..711Y} {362, 711}

\bibitem[\protect\citeauthoryear{Yang, Mo, van~den Bosch, Pasquali, Li  \&
  Barden}{Yang et~al.}{2007}]{Yang_2007}
Yang X.,  Mo H.~J.,  van~den Bosch F.~C.,  Pasquali A.,  Li C.,   Barden M.,
  2007, \mn@doi [\apj] {10.1086/522027}, 671, 153

\bibitem[\protect\citeauthoryear{{Yang}, {Zhu}, {Weijmans}, {van de Ven},
  {Boardman}, {Morganti}  \& {Oosterloo}}{{Yang} et~al.}{2020}]{Yang2020}
{Yang} M.,  {Zhu} L.,  {Weijmans} A.-M.,  {van de Ven} G.,  {Boardman} N.,
  {Morganti} R.,   {Oosterloo} T.,  2020, \mn@doi [\mnras]
  {10.1093/mnras/stz3293}, \href
  {https://ui.adsabs.harvard.edu/abs/2020MNRAS.491.4221Y} {491, 4221}

\bibitem[\protect\citeauthoryear{{Yao}, {Shan}, {Zhang}, {Kneib}  \&
  {Jullo}}{{Yao} et~al.}{2020}]{Yao2020}
{Yao} J.,  {Shan} H.,  {Zhang} P.,  {Kneib} J.-P.,   {Jullo} E.,  2020, \mn@doi
  [\apj] {10.3847/1538-4357/abc175}, \href
  {https://ui.adsabs.harvard.edu/abs/2020ApJ...904..135Y} {904, 135}

\bibitem[\protect\citeauthoryear{{Zhan}}{{Zhan}}{2011}]{zhanhu2011SSPMA..41.1441Z}
{Zhan} H.,  2011, \mn@doi [Scientia Sinica Physica, Mechanica \& Astronomica]
  {10.1360/132011-961}, \href
  {https://ui.adsabs.harvard.edu/abs/2011SSPMA..41.1441Z} {41, 1441}

\bibitem[\protect\citeauthoryear{{Zhan}}{{Zhan}}{2021}]{Zhanhu2021}
{Zhan} H.,  2021, \mn@doi [Chinese Science Bulletin] {10.1360/TB-2021-0016},
  66, 11

\bibitem[\protect\citeauthoryear{{Zhu}, {Lu}, {Cappellari}, {Li}, {Mao}  \&
  {Gao}}{{Zhu} et~al.}{2023a}]{zhukai_2023_paper2}
{Zhu} K.,  {Lu} S.,  {Cappellari} M.,  {Li} R.,  {Mao} S.,   {Gao} L.,  2023a,
  \mn@doi [arXiv e-prints] {10.48550/arXiv.2304.11714}, \href
  {https://ui.adsabs.harvard.edu/abs/2023arXiv230411714Z} {p. arXiv:2304.11714}

\bibitem[\protect\citeauthoryear{Zhu, Lu, Cappellari, Li, Mao  \& Gao}{Zhu
  et~al.}{2023b}]{Zhu_2023_paperI}
Zhu K.,  Lu S.,  Cappellari M.,  Li R.,  Mao S.,   Gao L.,  2023b, \mn@doi
  [\mnras] {10.1093/mnras/stad1299}, 522, 6326

\bibitem[\protect\citeauthoryear{Zou, Gao, Zhou  \& Kong}{Zou
  et~al.}{2019}]{Zou2019}
Zou H.,  Gao J.,  Zhou X.,   Kong X.,  2019, \mn@doi [\apjs]
  {10.3847/1538-4365/ab1847}, 242, 8

\bibitem[\protect\citeauthoryear{{Zu} et~al.,}{{Zu} et~al.}{2021}]{Zu2021}
{Zu} Y.,  et~al., 2021, \mn@doi [\mnras] {10.1093/mnras/stab1712}, \href
  {https://ui.adsabs.harvard.edu/abs/2021MNRAS.505.5117Z} {505, 5117}

\bibitem[\protect\citeauthoryear{{de Vaucouleurs}, {de Vaucouleurs}, {Corwin},
  {Buta}, {Paturel}  \& {Fouque}}{{de Vaucouleurs}
  et~al.}{1991}]{deVaucouleurs1991}
{de Vaucouleurs} G.,  {de Vaucouleurs} A.,  {Corwin} Herold~G. J.,  {Buta}
  R.~J.,  {Paturel} G.,   {Fouque} P.,  1991, {Third Reference Catalogue of
  Bright Galaxies}

\makeatother
\end{thebibliography}

	%\begin{thebibliography}{99}
	%\end{thebibliography}
	
	%%%%%%%%%%%%%%%%%%%%%%%%%%%%%%%%%%%%%%%%%%%%%%%%%%
	
	%%%%%%%%%%%%%%%%% APPENDICES %%%%%%%%%%%%%%%%%%%%%
	
	\appendix
	\section{}
	
\begin{figure}
		\centering
		\includegraphics[width=\columnwidth]{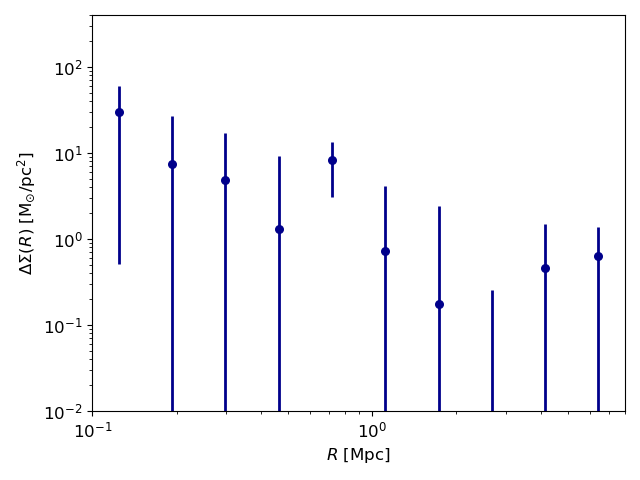}\\
		\caption{The gravitational lensing signal for the $M_{\rm 200m}<10^{13}{\rm M_{\odot}}$ sub-sample lens bin.}
		\label{fig:ggl_low_mass}
	\end{figure}

	We show the lensing signal together with measurement error bars in Fig.~\ref{fig:ggl_low_mass} for the halo mass bin $M_{\rm 200m}<10^{13}~{\rm [M_{\odot}])}$. Error bars are the 1$\sigma$ error. The S/N is too low to derive a meaningful model, thus we do not use the data of this mass bin in this paper.  

	%
	
	%%%%%%%%%%%%%%%%%%%%%%%%%%%%%%%%%%%%%%%%%%%%%%%%%%

	% Don't change these lines
	\bsp	% typesetting comment
	\label{lastpage}
\end{document}